\documentclass[aps,twocolumn,floatfix,showpacs,superscriptaddress]{revtex4}
\usepackage{amssymb,amsmath,graphicx,times}
\usepackage{mathptmx}
\usepackage[usenames]{color}

\usepackage{amsfonts,amsthm,bbm} 
\usepackage{physics}
\usepackage{mathtools}
\usepackage{lscape}
\usepackage{multirow}
\usepackage{diagbox}
\usepackage{color}
\usepackage[dvipsnames]{xcolor}
\usepackage{hyperref}
\usepackage{tensor}

\def\ket#1{{|#1\rangle}}
\def\bra#1{{\langle#1|}}
\def\ketbra#1{{|#1\rangle\langle#1|}}

\newtheorem{prop}{PROPOSITION}

\newtheorem{defi}{DEFINITION}

\begin{document}
\title{Iso-entangled mutually unbiased bases, symmetric
    quantum measurements  and mixed-state designs}
        
\author{Jakub Czartowski}
\affiliation{Institute of Physics, Jagiellonian University, ul. {\L}ojasiewicza 11, 30--348 Krak\'ow, Poland}

\author{Dardo Goyeneche}
\affiliation{Departamento de F\'{i}sica, 
Universidad de Antofagasta, 
Casilla 170, Antofagasta, Chile}

\author{Markus Grassl}
\affiliation{Max Planck Institute for the Science of Light, 91058 Erlangen, Germany}
  
\author{Karol {\.Z}yczkowski}
\affiliation{Institute of Physics, Jagiellonian University, ul. {\L}ojasiewicza 11, 30--348 Krak\'ow, Poland} 
\affiliation{Center for Theoretical Physics, Polish Academy of Sciences, Al. Lotnik\'{o}w 32/46, 02-668 Warszawa, Poland}

\pacs{03.67.-a, 03.65.Ud}

\date{November 3, 2019}

\begin{abstract}
\noindent
Discrete structures in Hilbert space play a crucial role in finding
optimal schemes for quantum measurements.  We solve the problem
whether a complete set of five iso-entangled mutually unbiased bases
exists in dimension four, providing an explicit analytical
construction.  The reduced density matrices of these $20$ pure states forming this
generalized quantum measurement form a regular dodecahedron inscribed
in a sphere of radius $\sqrt{3/20}$ located inside the Bloch ball of
radius $1/2$.  Such a set forms a mixed-state $2$-design --- a discrete
set of quantum states with the property that the mean value of any
quadratic function of density matrices is equal to the integral over
the entire set of mixed states with respect to the flat
Hilbert-Schmidt measure.  We 
establish necessary and sufficient conditions 
mixed-state designs need to satisfy and 
present  general methods to construct them.  
Furthermore, it is shown that partial traces of
a projective design in a composite Hilbert space form a mixed-state design,
while decoherence 
 of elements of a projective design 
yields a design in the classical probability simplex.
We identify a distinguished 
two-qubit
orthogonal basis 
such that  four reduced states
 are evenly distributed inside the Bloch ball and 
form a mixed-state $2$-design.
\end{abstract}

\maketitle

\emph{Introduction.}---Recent 
progress of the theory of quantum information
triggered renewed interest in foundations of quantum mechanics.
Problems related to measurements of an unknown quantum state
attract particular interest. 
The powerful technique of state tomography \cite{AJK04,BK10},
allowing  one to recover 
a density matrix,
can be considered as a
generalized quantum measurement, 
determined by a suitable set of pure quantum states of a fixed size $d$.
Notable examples include {\sl symmetric informationally complete} 
(SIC)  measurements 
 \cite{RBSC04,S06} 
consisting of $d^2$ pure states, which form 
a regular simplex inscribed inside the convex set 
 $\Omega_d \subset {\mathbb R}^{d^2-1}$ of density matrices of size $d$,
and complete sets of $(d+1)$ {\sl mutually unbiased bases} (MUBs)  \cite{WF89}
such that the overlap of any two vectors belonging to different bases is constant.

\emph{Introduction.}---Recent 
progress of the theory of quantum information
triggered renewed interest in foundations of quantum mechanics.
Problems related to measurements of an unknown quantum state
attract particular interest. 
The powerful technique of state tomography \cite{AJK04,BK10},
allowing  one to recover 
a density matrix,
can be considered as a
generalized quantum measurement, 
determined by a suitable set of pure quantum states of a fixed size $d$.
Notable examples include {\sl symmetric informationally complete} 
(SIC)  measurements 
 \cite{RBSC04,S06} 
consisting of $d^2$ pure states, which form 
a regular simplex inscribed inside the convex set 
 $\Omega_d \subset {\mathbb R}^{d^2-1}$ of density matrices of size $d$,
and complete sets of $(d+1)$ {\sl mutually unbiased bases} (MUBs)  \cite{WF89}
such that the overlap of any two vectors belonging to different bases is constant.

The above schemes are distinguished by the fact that they allow to
maximize the information obtained from a measurement and minimize the
uncertainty of the results obtained under the presence of errors in
both state preparation and measurement stages \cite{S06,AS10}.
Interestingly, it is still unknown, whether these configurations exist
for an arbitrary dimension.  In the case of SIC measurements
analytical results were known in some dimensions up to $d=48$, see
\cite{SG10} and references therein. More recently, a putative infinite
family of SICs starting with dimensions $d =4,8,19,48,124,323$ has been
constructed \cite{GS17}, while the general problem remains
open. Nonetheless, numerical results suggest \cite{SG10}
 that such configurations might
exist in every finite dimension $d$.  For MUBs explicit
constructions are known in every prime power dimension $d$
\cite{WF89}, and it is uncertain whether such a solution exists
otherwise, in particular \cite{BBELTZ07,DEBZ10} in dimension $d=6$.

If the dimension is a square, $d=N^2$, the system can be considered
as two subsystems of size $N$ and the effects of quantum entanglement
become relevant. It is possible to prove that the average entanglement
of all bi-partite states forming a SIC or a complete set of MUBs is
fixed \cite{L78}.

It is natural to ask whether there exists a particular configuration
such that all the states forming the generalized measurements 
 share the same amount of entanglement 
so that they are locally equivalent,
$|\phi'\rangle=U_A \otimes U_B |\phi\rangle$.
In the simplest case of 
$d=4$ 
 a set of $16$ iso-entangled vectors
forming a SIC was analytically constructed by Zhu, Teo and Englert \cite{ZTE10},
thus such a set can be obtained from a selected {\sl fiducial} state $|\phi\rangle$
by local unitary operations. Further entanglement properties of
SICs were studied in  \cite{LLZZ18,CGZ18}.
Although entanglement of the states forming MUBs in composite dimensions
was analyzed  
 \cite{La04,RBKS05,WPZ11,GG15}, 
the analogous problem of finding a full set of iso-entangled MUBs 
remained open till now even for two-qubit system.
 
Collections of states forming a SIC measurement or a set of MUBs find
numerous applications in the theory of quantum information
\cite{Re05,ZTE10,AS10,Ta12}.  They belong to the class of {\sl
  projective designs}: finite sets of evenly distributed pure quantum
states in a  given dimension $d$ such that the mean value of any function
from a certain class is equal to the integral over the set 
of pure states with respect to the unitarily invariant Fubini-Study
measure \cite{DGS77,RBSC04,BZ17}.  These discrete sets of pure quantum
states, and analogous sets of unitary operators called {\sl unitary
  designs} \cite{GAE07}, proved to be useful for process tomography
\cite{Sc08}, construction of unitary codes \cite{RS09}, realization of
quantum information protocols \cite{DCEL09}, derandomization of
probabilistic constructions \cite {GKK15}, and detection of
entanglement \cite{BHM18}.
 
A cognate notion of  quantum conical design was 
recently proposed
\cite{GA16, Szymusiak},
which concerns operators of an arbitrary rank from
the cone of mixed quantum states.
However, these designs are not suitable to
sample the set  $\Omega_d$ of mixed states   
according to the flat, Hilbert-Schmidt  measure. 
On the other hand the general theory of
{\sl averaging sets} developed in \cite{SZ84}
implies that such configurations 
of mixed quantum states do exist. 
 
In this letter
we solve the longstanding problem of existence of iso-entangled
MUBs in dimension four. Secondly, we introduce the 
notion of a quantum mixed-state design,
such that mean values of selected functions
over this discrete set of density matrices
equals to the average value integrated over the set $\Omega_d$,
and provide a notable example with dodecahedral symmetry
constructed from the constellation of iso-entangled MUBs.
Furthermore, we show that a projective $t$-design induces
by the coarse graining map 
a $t$-design in the classical probability simplex,
and establish general links between the designs in the
sets of classical and quantum states.

\bigskip
\emph{MUBs for bi-partite systems.}---The standard solution of $5$
MUBs in dimension $d=4$ consists of $12$ separable states forming
three bases and $8$ maximally entangled states corresponding
to the remaining two bases \cite{RBKS05,AS10}.
Thus the partial trace of these states yields a peculiar configuration
inside the Bloch ball: $6$ corners of a regular octahedron inscribed
into the Bloch sphere, covered by two points each, correspond to $3$
MUBs in ${\cal H}_2$.  The other $8$ points sit degenerated at the
center of the ball representing the maximally mixed state,   
${\mathbb I}/2$.
The total configuration
consists thus of $7$ points, at the expense of weighing the central
point as four points at the surface.
 Note that the Schmidt vectors of the
first twelve pure product states are $\lambda_{\rm sep}=(1,0)$, while
for the other eight states this vector reads $\lambda_{\rm ent}=(1/2,
1/2)$.  As this set of vectors in ${\cal H}_4$ forms a projective
$2$-design, the average degree of entanglement measured by purity is
fixed, $ \langle \lambda_1^2+\lambda_2^2 \rangle=4/5$.
For any dimension being a power of a prime, $d=p^k$, the standard
solution of the MUB problem consist of $p+1$ separable bases and 
$p^k-p$ maximally entangled bases \cite{La11}.  
In the
case of $d=9$ the set of MUBs consisting of $4$ separable and $6$
maximally entangled bases was studied by Lawrence \cite{La04}.

\smallskip
\emph{Two--qubit iso-entangled MUBs.}---As the set of iso-entangled
vectors forming a SIC is known for two \cite{ZTE10} and three
\cite{H82} qubit systems, it is natural to ask 
whether there exists an
analogous configuration of iso-entangled MUBs. In other words, we wish
to find a global unitary rotation $U\in U(4)$ acting on the 
standard 
constellation in such a way that the degeneracy of the configuration
of $20$ points is lifted and all of them
become equally distant from the center of the Bloch ball. 
Then the
corresponding vectors in ${\cal H}_4$ share the same degree of
entanglement and can be obtained from a selected {\sl fiducial vector}
$|\phi_1\rangle$ by local unitaries, $|\phi_j\rangle = U_j \otimes W_j
|\phi_1\rangle$ with $j=2,\dots, 20$.

We construct the desired set of five
iso-entangled MUBs in ${\cal H}_4$ making use of the fact that the
group of local unitary operations is in this case isomorphic to the
double cover of the alternating group $A_5$.  It has two faithful
irreducible representations of degree two and it admits a tensor
product representation that allows us to construct the necessary local
two-qubit gates $U_j \otimes W_j$.

\begin{figure}[h]
    \centering
    \includegraphics[width=.8\linewidth]{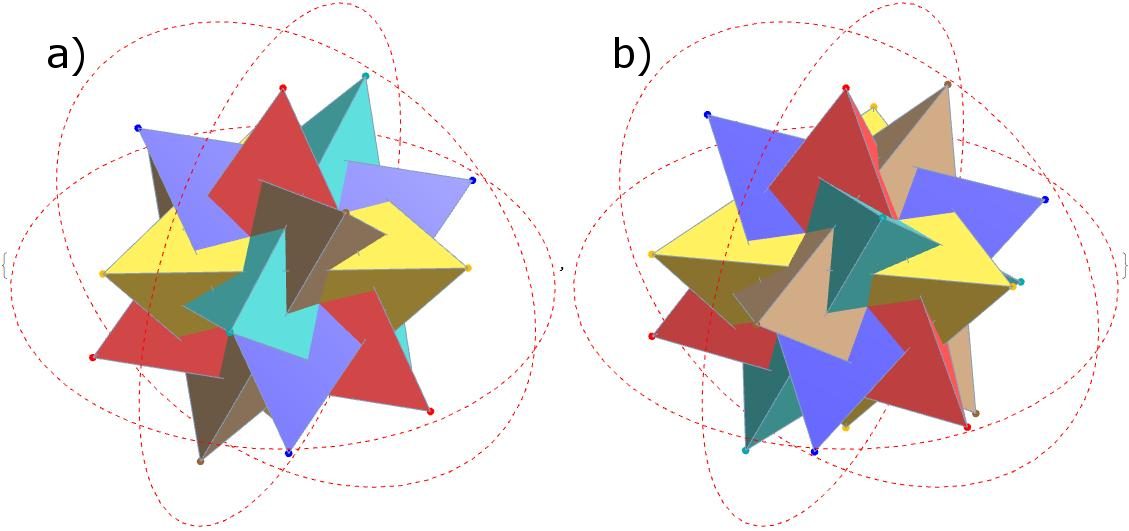}
    \caption{One-qubit mixed-state design composed of $20$ points
      inside the Bloch ball of radius $1/2$ obtained by partial trace of
      the $20$ states in $\mathcal{H}_2 \otimes \mathcal{H}_2$, which
      form a set of iso-entangled mutually unbiased bases for two
      qubits.  Each basis is represented by the vertices of a regular
      tetrahedron inscribed in the sphere of radius $r=\sqrt{3/20}$.
      The reduced density matrices on both subsystems are shown in a)
      and b).}
    \label{2221}
\end{figure}

As shown in Appendix \ref{20MUB},
the full analytic solution can be generated by local unitaries
from the following fiducial state,
\begin{equation} 
\label{fiducial_vector}
	\ket{\phi_1} = \frac{1}{20}\qty(a_+\ket{00} - 10i\ket{01} + \qty(8i-6)\ket{10} + a_- \ket{11}),
\end{equation}
where $a_\pm = -7 \pm 3\sqrt{5}  + i(1 \pm\sqrt{5})$. 
Since the states forming five bases are iso-entangled, their
partial traces with respect to the first (or the second) subsystem
share the same purity and 
belong to a sphere of radius $r=\sqrt{3/20}$, 
 embedded inside the Bloch ball of radius $R=1/2$.  
 The set of $20$ points enjoys a dodecahedral symmetry,
shown in Fig.  \ref{2221}.  Reductions of the four states stemming
from each of the five bases in $\mathcal{H}_4$ form a regular
tetrahedron in both reductions, so up to rescaling their Bloch vectors
form a SIC for a single qubit.  In both reductions the mixed states corresponding to all five
bases form a five-tetrahedron compound with the
same chirality, while their convex hull yields a regular dodecahedron.
This  
configuration is not directly related to the arrangement of $20$
pure states in dimension $4$ forming the {\sl magic dodecahedron} of
Penrose \cite{Pe94,ZP93,MA99}.  It differs also from
the regular dodecahedron of Zimba \cite{Zi06}, which describes a
basis of five orthogonal anticoherent states in ${\cal H}_5$ in the
stellar representation.

\emph{Projective and unitary designs.}---Recall that
a {\sl projective $t$-design} is an 
ensemble  of $M$  pure states,  $\qty{\ket{\psi_j}\in {\cal H}_d}_{j=1}^M$,
such that
for any polynomial $f_t$  of the state $\psi $ of degree at most $t$
 its average value is equal to the integral with respect to the unitarily invariant
Fubini--Study measure  $\dd{\psi}_{FS}$
over the entire  complex projective space of pure states,
$\Xi_d={\mathbbm C}P^{\; d-1}$,
\begin{equation}\label{cptd}
    \frac{1}{M}  \sum_{j=1}^{M} f_t\qty(\psi_j) = \int_{\Xi_d} f_t\qty(\psi) \dd{\psi}_{FS}. 
\end{equation}
The notions of pure-state $t$-designs and unitary $t$-designs,
consisting of matrices evenly distributed over the unitary group \cite{GAE07},
found numerous applications in quantum information processing
 \cite{Sc08,RS09,DCEL09,GKK15}
  and have been applied in experiments \cite{Ta12, BQTSLSKB15, BHM18}.
 They can be considered as a special case of 
 {\sl averaging sets}, which are known to exist for
 arbitrary sets endowed with a probability measure \cite{SZ84}.
Below we shall adopt this notion to the set of density matrices
and show how such mixed-state designs can be constructed. 

\smallskip
\emph{Mixed-state designs.}---We shall start by introducing a formal definition
of mixed-state $t$-designs  with respect to the Hilbert-Schmidt measure
in the space of density matrices.

\begin{defi}
    A collection of $M$ density matrices $\{\rho_i \in \Omega_N\}_{i=1}^M$ is called a \textbf{mixed-state t-design} if for any polynomial $g_t$ 
     of the state $\rho$  of degree $t$
    the average over the collection is equal to the mean value over the
    set  $\Omega_N$ of mixed states in dimension $N$ with respect to the normalized Hilbert-Schmidt measure $\dd{\rho_{\rm HS}}$,
    \begin{equation}
        \frac{1}{M} \sum_{i=1}^{M} g_t \qty(\rho_i) = \int_{\Omega_N} g_t\qty(\rho) \dd{\rho_{\rm HS}}. 
        \label{defi1}
    \end{equation} 
\end{defi}
The above condition,  analogous to the definition of  projective $t$-designs (\ref{cptd}),
is equivalent to the following relation,
\begin{equation}
    \frac{1}{M}\sum_{i=1}^M \rho_i^{\otimes t} =
    \int_{\Omega_N} \rho^{\otimes t}\dd{\rho_{\rm HS}} =:   \omega_{N,t},
    \label{omega}
\end{equation}
where the mean product state of a system consisting of $t$ copies 
of a state $\rho$ in dimension $N$ averaged 
over the entire space  ${\Omega_N}$ of mixed states
 is denoted by $\omega_{N,t}$.
The measure $\dd{\rho_{\rm HS}}$ is defined by the  requirement that each unit ball 
with respect to the Hilbert-Schmidt distance has the same volume.
     
Observe that for $t=1$ Definition \eqref{defi1} reduces  to 
a resolution of the maximally mixed state,
$
    \frac{1}{M}\sum_{i=1}^M \rho_i = \frac{1}{N}{\mathbb{I}_N}
$
so any mixed-state design forms a generalized quantum
measurement (also called POVM). To check whether a given configuration of
density matrices forms a $t$-design we establish the following 
 necessary and sufficient condition.
\begin{prop}\label{prop1}  
    A set consisting of $M$ states from the set $\Omega_N$ of density matrices of size $N$
    forms a mixed-state $t$-design if and only if the following bound is saturated,
    \begin{equation}
        2 \Tr(  \omega_{N,t} \; \frac{1}{M} \sum_{i=1}^M \rho_i^{\otimes t} 
       ) -
        \frac{1}{M^2}\sum_{i,j=1}^M \Tr(\rho_i\rho_j)^t
        \leq
        \gamma_{N,t}
        \label{welchprop1}
    \end{equation}
    where $\gamma_{N,t} := \Tr \omega^2_{N,t}$
    with $\omega_{N,t}$ defined by Eq. (\ref{omega}).
\end{prop}

This condition, proved in Appendix \ref{APP_proof_1}
is closely related to saturation of the
Welch bound \cite{We74}
for projective and unitary designs \cite{Sc08}.
Such an tool allows one to construct  such designs by numerical minimization. 
Exact values of $\gamma_{N,t}$ for $t\leq5$ are given in Appendix \ref{APP_general_N}.

Using the bound \eqref{welchprop1}, we were able to find numerical
lower bounds for the number $M$ of states in a mixed $2$-design:
$M\ge 4$  for $N=2$ and $M\ge 9$ for $N=3$.  In particular,
for $N=2$ the  minimal mixed-state $2$-design forms a
tetrahedron inside the Bloch ball, an example of Platonic designs,
equivalent to a single tetrahedron out of five plotted in Fig.  \ref{2221} -- see Appendix \ref{platonic_des}.

\emph{Connection between pure- and mixed-state designs.}---We will
show that a mixed-state design for a single system of size
$N$ can be generated from a bipartite pure-state 
design of size $N\times N$.
Since such constellations exist for all dimensions, the following
result, proved in Appendix \ref{App_prop2}, implies that mixed-state
$t$-designs exist for every $N$.

\begin{prop}
\label{prop2}
    Any complex projective $s$-design
    $\qty{\ket{\psi_j}}_{j=1}^M$ in the composite Hilbert space
    $\mathcal{H}_{A}\otimes\mathcal{H}_{B}$ of dimension $d=N^2$
    induces by partial trace a mixed-state $t$-design $\qty{\rho_{j}}_{j=1}^M$ in
    $\Omega_N$ with  $t\ge s$ and $\rho_{j} = \Tr_B\ketbra{\psi_j}$.
     The same property holds also for the dual set $\{ \rho'_{j} =
    \Tr_A\ketbra{\psi_j}\}_{j=1}^M$.
\end{prop}
 
In particular, Proposition \ref{prop2} implies that taking partial
trace of pure states forming a SIC 
 in ${\cal H}_{N^2}$, or any other pure state  $2$-design,
 one obtains a mixed-state
$2$-design in the set $\Omega_N$ of density matrices of size $N$.
Interestingly, there exist distinguished cases for which the degree 
of  the design increases, $t>s$:
In Appendix \ref{examples} 
we demonstrate that partial trace of any orthogonal basis, $t=1$, 
of the five iso-entangled MUBs
yields a mixed state $2$-design,
while the complete set of these MUBs, $t=2$, 
leads to  a mixed state $3$-design.
Furthermore, the following one-to-one
relation between a class of mixed-state $2$-designs
and projective $2$-designs is proven in Appendix \ref{app_dilut}.
\begin{prop} \label{dilution_prop}
	Any projective 2-design $\qty{\ket{\psi_i}}_{i=1}^N$ of dimension $N$ can be diluted into a mixed 2-design by taking 
	projectors onto all states forming the projective 2-design with weight $p = \frac{1+N}{1+N^2}$ and the maximally mixed state $\mathbb{I}_N/N$
	 with weight $1-p = \frac{N^2 - N}{1 + N^2}$.
\end{prop}
%

\medskip

\emph{Designs in classical probability simplex.
}---To construct one-qubit mixed-state designs 
one needs to determine the radial distribution of points inside the Bloch ball. 
It is related to an averaging set on the interval $[-1/2, 1/2]$ 
with respect to the Hilbert--Schmidt (HS) measure \cite{ZS01}
 determining the distribution
of eigenvalues of a random mixed quantum state.

Returning to the general case of an arbitrary dimension $N$,
consider any fixed probability measure $\mu\qty(x)$ defined on
the simplex $\Delta_{N}$ of $N$-point probability vectors.
We wish to find an averaging set over the simplex, i.e., a sequence
of $M$ points $\qty{x_i\colon x_i\in \Delta_N}_{i=1}^M$ which satisfy
the condition analogous to $t$-designs, with respect to the integration
measure $\mu(x)$:
    \begin{equation}
        \frac{1}{M}\sum_{i=1}^M f_t(x_i) = \int_{\Delta_N} f_t(x) \mu\qty(x)\dd{x},
        \label{eigtdes}
    \end{equation}
where $f_t$ denotes an arbitrary polynomial of order $t$.

\begin{figure}
    \centering
    \includegraphics[width=.8\linewidth]{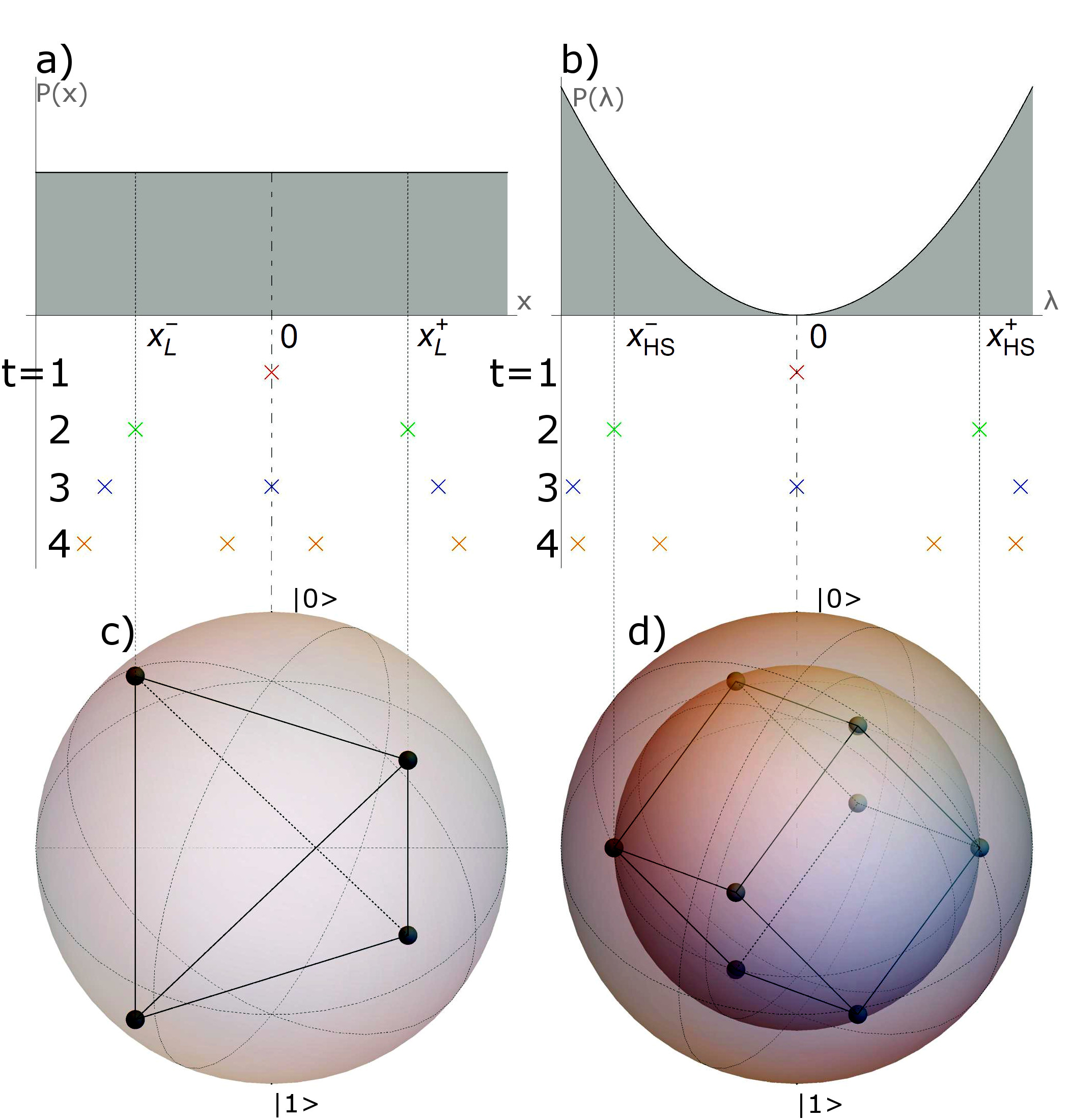}
    \caption{Simplicial $t$-designs  on $\Delta_2=[-1/2,1/2]$ for $t=1,2,3,4$
     with respect to a) flat measure and b) Hilbert-Schmidt measure. c) $2$-design with respect to the flat measure
    $\mu_{\rm L}$  corresponds to 
    the $x$  coordinates of a tetrahedron inscribed in a Bloch sphere,
  related to one-qubit projective $2$-design produced by a SIC-POVM in ${\cal H}_2$.
   d) $2$-design with respect to the HS measure corresponds to the radius of the	 sphere containing
    the mixed-state $2$-design -- 
   the cube induced by the iso-entangled SIC-POVM in ${\cal H}_4$.} 
    \label{fig:my_label}
\end{figure}

Exemplary minimal solutions 
of this problem for low values of $t$ and $N=2$,
so that the integration is done over the interval $\Delta_2=[-1/2,
  1/2]$, are presented in Appendix \ref{APP_interval}.  
Here we shall
concentrate on the cases of $t=1,2$ for the Lebesgue and HS measure,
as these results are linked to one-qubit pure and mixed-state
designs, respectively.  1-design in the interval with respect to
both measures consists of a single point in its center, corresponding
to the projection on the $x$ axis of the basis $|0\rangle, |1\rangle$,
which yields both projective and mixed-state $1$-design.  Interval
$2$-design with respect the flat Lebesgue measure, $\mu_{\rm L}(x)=1$,
gives coordinates of vertices of a tetrahedron inscribed in a sphere of
unit radius, i.e., a SIC-POVM in dimension $2$. The analogous design with
respect to $\mu_{\rm HS}$ 
provides the radius of a sphere in the Bloch ball
containing mixed-state 2-designs with constant purity.
An exemplary 2-design obtained by partial trace of 16 states
forming an iso-entangled SIC-POVM for 2 qubits
is shown in Fig. \ref{fig:my_label}d.
   
  Positions of both points at the unit interval,
  which form $2$-designs with respect to both measures,
   $ x_{\pm}^{\rm L} = \pm 1/2\sqrt{3}$
   and 
   $ x_{\pm}^{\rm HS} = \pm \sqrt{3/20}$,
   can be thus related to the geometry of regular bodies inscribed into
   a sphere.  Note that the design on $[0, 1]$ with respect 
   to the flat measure is formed by probabilities $p_i=|\langle i|\psi_j\rangle|^2$
   related to projections of the states of the design 
    onto the computational basis.  This observation,
   corresponding to the decoherence of a quantum state
   to the classical probability vector,
   can be generalized for higher dimensions.
\begin{prop}
\label{prop3}
    Any complex projective $t$-design $\qty{\ket{\psi_j}}_{j=1}^M$ in the Hilbert space 
    $\mathcal{H}_N$ induces, by the coarse graining map, 
      $|\psi\rangle \langle \psi|\to {\vec p}:={\rm diag}(|\psi\rangle \langle \psi|)$,
       a $t$-design in the $N$--point classical probability simplex $\Delta_{N}$
       with respect to the flat measure $\mu_{\rm L}$.
\end{prop}
To prove this fact it is sufficient to recall that the natural,
unitarily invariant measure in the space of pure states induces, by
decoherence, the flat measure $\mu_{\rm L}$ in the probability
simplex, see Appendix \ref{APP_interval}.
The notion of $t$-designs formulated for a
probability simplex allows one to select classical states which are
useful to approximate an integral over the entire set $\Delta_{N}$. 
This also implies a simple, yet important observation that
 a mixed-state design in  dimension $N=2$ with $t>3$
 cannot be generated from iso-entangled pure states in ${\cal H}_4$.

Furthermore, we suggest a general approach to obtain
mixed designs of a product form. 
It will be convenient to use an asymmetric
part $\tilde \Delta_N$ of the simplex $\Delta_N$,
which corresponds  to ordering of eigenvalues, 
$\lambda_1 \ge \lambda_2 \ge \dots \ge \lambda_N$.
\begin{prop}
\label{prop4}
Consider a $t$-design $\qty{\lambda_i}_{i=1}^n$ in the simplex
$\Delta_N$ with respect to the measure $\mu_{\rm HS}$, the
corresponding set of diagonal matrices $\Lambda_i = {\rm
  diag}\qty(\lambda_i)$ and any unitary $t$-design
$\qty{U_j}_{j=1}^m$.  Let $n'$ denote the number of points of the
simplicial design belonging to the asymmetric part $\tilde \Delta_N$.
Then the Cartesian product consisting of $n'm$ density matrices,
$\rho_{ij}=U_j\Lambda_i U_j^{\dagger}$, $i=1,\dots, n'$ and $j=1,
\dots, m$, forms a mixed-state $t$-design in $\Omega_N$.
\end{prop}

This statement, demonstrated in Appendix \ref{App_prop5},
allows us to construct Platonic mixed-state $t$-designs inside the Bloch ball:
restricting the HS $2$-design in $\Delta_2$ to its half $\tilde
\Delta_2=[0,1/2]$ we arrive at a single point $x_{+}^{\rm HS}
=\sqrt{3/20}$, which determines the radius of the sphere inside the
Bloch ball.  Taking the corresponding spectrum, $\Lambda={\rm
  diag}(1/2+ x_{+}^{\rm HS},1/2- x_{+}^{\rm HS})$, and rotating it by
unitaries $U_i$ from a unitary design in $SU(2)$ we arrive at a
mixed-state design. In the simplest case of the tetrahedral group
the mixed-state $2$-design consists of four points
forming one of the five tetrahedrons shown in Fig. \ref{2221},
which arise  by partial trace of the iso-entangled bases 
listed in Appendix \ref{platonic_des}.
This example shows that there exist mixed state $t$-designs
which cannot be purified to a pure state $t$-design.

\bigskip
\emph{Outlook and conclusions.}---
In this work we introduced the notion of  mixed-state
$t$-designs and established 
 necessary and sufficient conditions for their existence.
As any mixed-state $1$-design forms a POVM, any  design of a
higher order $t$ can be considered as a generalized measurement with
additional symmetry properties \cite{Zh15}.
From the physical perspective such a
 deterministic sequence of density matrices
 approximates a sample of random states
 and  describes projective designs on a bipartite system AB,
under the restriction that Alice 
receives no information from Bob. 

Analyzing mixed-states designs we solved the problem of 
existence of $20$ locally equivalent two-qubit states which form a set of five MUBs. 
 The obtained configuration defines a
remarkable  
 measurement scheme,  
useful for quantum state estimation \cite{RTH15} and for constructing
symmetric entanglement witnesses based on MUBs \cite{SHBAH12, CSW18},
different  from those analyzed
earlier \cite{MBPKIN14,BCLM16}. 
 We analytically derived a two-qubit fiducial state,
so that the other states forming the five bases 
were obtained by applying local unitaries. 
The partial trace of these  two-qubit states
forms a structure with dodecahedral symmetry
 inscribed into a sphere inside the Bloch ball. 
 This particular configuration consisting of five tetrahedrons,
 visualized in Fig. \ref{2221}, 
 leads to a notable example of a mixed-state $3$-design. 
Each single tetrahedron, obtained by partial trace of a single basis,
forms a $2$-design.

The paper establishes a direct link between designs in various sets
which serve as a scene for quantum information processing: any
projective $t$-design composed of pure states in dimension $d=N^2$
 induces by partial trace a
mixed-state design in the set of density matrices in dimension $N$,
while by the decoherence channel 
 it produces a design in the classical $d$-point probability simplex. 
A class of mixed-state designs can be constructed by the Cartesian product of 
a unitary design and a simplicial Hilbert-Schmidt design.
These relations, based on transformations of measures, put the notion of
designs in various spaces into a common framework, and show how to
approximate averaging over continuous sets by discrete sums. Such an
approach is not only of direct interest for 
theoretical work on   
 foundations of quantum mechanics, but also
for experimental realization of an approximate 
ensemble of random quantum or classical states.

We shall conclude the paper with a brief list of  open problems:
 \emph{(i)} Find the minimal number of elements $M(N)$ forming a
 minimal mixed $t$-design in dimension $N$;
 \emph{(ii)} Find minimal mixed-state  $t$-designs,
  for which the variance of the purity of all the states
 is the smallest; \emph{(iii)} 
 Numerical calculations performed for $N=3,4,5 $  suggest
 that there exist orthogonal  bases in ${\cal H}_N \otimes {\cal H}_N$
 such that their partial trace gives a mixed state $2$-design in $\Omega_N$.
  Determine, whether this conjecture, 
  proved  here for $N=2$,  holds also for higher dimensions.
  %

\bigskip
\emph{Acknowledgements.}---It is a pleasure to thank M. Appleby, 
K. Bartkiewicz, I. Bengtsson, B.C. Hiesmayr,  P. Horodecki,
{\L}. Rudnicki and A. Szymusiak for inspiring discussions. 
Financial support by Narodowe Centrum Nauki 
under the grant number DEC-2015/18/A/ST2/00274 
and by Foundation for Polish Science 
under the Team-Net project is gratefully acknowledged. 
DG is supported by MINEDUC-UA project, code ANT 1855 
and Grant FONDECYT Iniciaci\'{o}n number 11180474, Chile.

\begin{widetext}

\appendix

\section{Explicit form of $20$ iso-entangled states forming $5$ MUBs}
\label{20MUB}

The standard construction of  a complete set of two-qubit mutually unbiased bases
using finite fields yields the following five bases $\qty{\ket{\psi_i}}_{i=1}^{20}$,
written row-wise, with normalization omitted  \cite{RBKS05},
\begin{alignat}{10}
&\ket{00}&
&\ket{01}&
&\ket{10}&
&\ket{11}\label{eq:MUB1}
\\
&(\ket{0}+\ket{1})(\ket{0}+\ket{1})&
&(\ket{0}+\ket{1})(\ket{0}-\ket{1})&
&(\ket{0}-\ket{1})(\ket{0}+\ket{1})&
&(\ket{0}-\ket{1})(\ket{0}-\ket{1})\label{eq:MUB2}
\\
&(\ket{0}+i\ket{1})(\ket{0}+i\ket{1})&
&(\ket{0}+i\ket{1})(\ket{0}-i\ket{1})&
&(\ket{0}-i\ket{1})(\ket{0}+i\ket{1})&
&(\ket{0}-i\ket{1})(\ket{0}-i\ket{1})\label{eq:MUB3}
\\
&\ket{00}-\ket{10}+i\ket{01}+i\ket{11}&\quad
&\ket{00}-\ket{10}-i\ket{01}-i\ket{11}&\quad
&\ket{00}+\ket{10}-i\ket{01}+i\ket{11}&\quad
&\ket{00}+\ket{10}+i\ket{01}-i\ket{11}\label{eq:MUB4}
\\
&\ket{00}-i\ket{10}+\ket{01}+i\ket{11}&\quad
&\ket{00}+i\ket{10}+\ket{01}-i\ket{11}&\quad
&\ket{00}-i\ket{10}-\ket{01}-i\ket{11}&\quad
&\ket{00}+i\ket{10}-\ket{01}+i\ket{11}.
\label{eq:MUB5}
\end{alignat}
The first three bases consist of product vectors, while the states in the last
two bases are all maximally entangled,
as the corresponding matrices of coefficients are unitary.
  The group $G_{\text{sym}}$ of
unitary matrices that map the set of $20$ vectors onto itself up to
phases, $G \ket{\psi_k} = e^{i \chi_{jk}}\ket{\psi_j}$,
 is generated by two complex Hadamard matrices,
(One can always add multiples of identity $e^{i\phi}I$
  to the group, but we consider the smallest possible group here),
  
\begin{alignat}{5}
  G_{\text{sym}}=\left\langle
  \frac{1}{2}
    \begin{pmatrix}
      i & -i &  1 &  1\\
      i & -i & -1 & -1\\
     -i & -i & -1 &  1\\
     -i & -i &  1 & -1
    \end{pmatrix},
  \frac{1}{2}
    \begin{pmatrix}
     -i &  i & -i &  i\\
      1 &  1 & -1 & -1\\
      1 &  1 &  1 &  1\\
      i & -i & -i &  i
    \end{pmatrix}
  \right\rangle.
\end{alignat}
The group is a subgroup of the so-called Clifford group that maps
tensor products of Pauli matrices onto itself. The group
$G_{\text{sym}}$ has order $7680$, and its center is generated by
$iI$, i.e., it has order four. The action on the $20$ states modulo
phases is a permutation group $G_{\text{perm}}$ of order $1920$. The
group acts transitively, i.e., any state can be mapped to any other
state.

Assume that we can find a subgroup $H\le G_{\text{sym}}$ that acts
transitively on the $20$ states and that, after a global change of
basis, all elements of $H$ can be written as tensor products.  In the
transformed basis, we will then obtain a complete set of MUBs such
that all the states forming the bases are equivalent up to local
unitaries, so they share the same Schmidt vector.  Unfortunately, the
problem of deciding whether a finite matrix group can be expressed as
a tensor product appears to be non-trivial in general.  There are both
necessary and sufficient conditions, but there does not seem to be a
simple general criterion.

In our case, there are transitive subgroups of $G_{\text{perm}}$ of
order  $20$, $60$, $80$, $120$, $160$, $320$, $960$, and $1920$. By
direct solving the equations for a change of basis that transforms all
elements of the corresponding matrix group into tensor products, we
find that only the subgroup $H_{\text{perm}}$ of order $60$ affords a
representation as a tensor product. The group $H_{\text{perm}}$ is
isomorphic to the alternating group $A_5$ on five letters. The
corresponding subgroup $H_{\text{sym}}$ is generated by
\begin{alignat}{5}\label{eq:gens_H_sym}
  H_{\text{sym}}=\left\langle
  \frac{1}{2}
    \begin{pmatrix}
      -1 &  1 & -i & -i\\
       1 & -1 & -i & -i\\
       i &  i &  1 & -1\\
       i &  i & -1 &  1
    \end{pmatrix},
  \frac{1}{2}
    \begin{pmatrix}
       i &  i &  i &  i\\
      -1 &  1 & -1 &  1\\
      -1 & -1 &  1 &  1\\
      -i &  i &  i & -i
    \end{pmatrix}
  \right\rangle.
\end{alignat}
The group $H_{\text{sym}}$ is also isomorphic to $A_5$, and its center
is trivial.  The group $A_5$ does not have a faithful representation of
degree $2$, and hence $H_{\text{sym}}$ cannot be written as the tensor
product of two representations of $A_5$.  However, the double cover of
$A_5$, which is isomorphic to the group $\mathrm{SL}(2,5)$ of $2\times 2$
matrices over the integers modulo $5$ with determinant $1$, has two
faithful irreducible representations of degree $2$.  The tensor
product of these two representation yields a group of order $60$ that
is conjugate to $H_{\text{sym}}$.

A global change of basis that transforms $H_{\text{sym}}$ into a
tensor product is given by
\begin{alignat}{5}
  H_{\text{local}}= T H_{\text{sym}} T^\dag,
\end{alignat}
where the global unitary transform reads
\begin{small}
\begin{alignat}{5}
  \def\w{\zeta}
  T=\frac{1}{20}\begin{pmatrix}
 -(\sqrt{5} + 1)i - 3\sqrt{5} + 7 & (\sqrt{5} - 1)i + 3\sqrt{5} + 7 & (\sqrt{5} - 1)i + 3\sqrt{5} + 7 & (-3\sqrt{5} + 7)i + \sqrt{5} + 1 \\
 10i & 10 & -10 & 10 \\
 -8i + 6 & 6i + 8 & -6i - 8 & -6i - 8 \\
 (\sqrt{5} - 1)i + 3\sqrt{5} + 7 & -(\sqrt{5} + 1)i - 3\sqrt{5} + 7 & -(\sqrt{5} + 1)i - 3\sqrt{5} + 7 & (3\sqrt{5} + 7)i - \sqrt{5} + 1
  \end{pmatrix}.
\end{alignat}
\end{small}%
Explicitly, we obtain two local generators $h_1, h_2$ (which do not
directly correspond to those in \eqref{eq:gens_H_sym})
\begin{small}
\begin{alignat}{5}
  H_{\text{local}}=\Biggl\langle
& \frac{1}{50}\begin{pmatrix}
    5 & -2i\sqrt{5} + \sqrt{5}\\
    i\sqrt{5} - 2\sqrt{5} & -5i
  \end{pmatrix}
  \otimes
  \begin{pmatrix}
     5                      & 2i\sqrt{5} + \sqrt{5}\\
     -i\sqrt{5} - 2\sqrt{5} & 5i
  \end{pmatrix},\\
& \frac{1}{20^2}\begin{pmatrix}
    (5\sqrt{5} + 5)i & (4\sqrt{5} - 10)i + 3\sqrt{5} + 5\\
    (4\sqrt{5} - 10)i - 3\sqrt{5} - 5 & -(5\sqrt{5} + 5)i
  \end{pmatrix}
  \otimes
   \begin{pmatrix}
     (5\sqrt{5} - 5)i                   & -(4\sqrt{5} + 10)i + 3\sqrt{5} - 5\\ 
     -(4\sqrt{5} + 10)i - 3\sqrt{5} + 5 & (-5\sqrt{5} + 5)i
   \end{pmatrix}
  \Biggr\rangle.
\end{alignat}
\end{small}%
In this basis, we see that the first and the second tensor factor are
similar, but not identical; they correspond to inequivalent
representations of $\mathrm{SL}(2,5)$.  Applying the transformation
$T$ to the complete MUB in \eqref{eq:MUB1}--\eqref{eq:MUB5} we obtain
the $20$ states of the iso-entangled complete set of MUBs shown in
Table \ref{table:isoMUB}. Partial traces over both subsystems of these 20 states form regular dodecahedra in the Bloch ball, shown in Fig. 2. Both configurations are related by an antiunitary transformation, which includes multiplication by a diagonal matrix with diagonal $\qty(1,\,i)$ and complex conjugation. The phases are chosen such that the action
of $H_{\text{local}}$ on these states does not introduce additional
phase factors. 

\vfill

\begin{table}[hbt]
\begin{alignat*}{10}\def\arraystretch{1.4}
\begin{array}{|c|c|c|c|c|}
\hline
\text{group element}&  \ket{00} &\ket{01} & \ket{10} & \ket{11}\\
\hline
id                              & (\sqrt{5} + 1)i + 3\sqrt{5} - 7 & -10i & 8i - 6 & (-\sqrt{5} + 1)i - 3\sqrt{5} - 7 \\
h_2h_1^2h_2h_1h_2               & (\sqrt{5} + 1)i + 3\sqrt{5} - 7 & 10i & -8i + 6 & (-\sqrt{5} + 1)i - 3\sqrt{5} - 7 \\
h_2h_1h_2h_1h_2h_1^2h_2         & (-\sqrt{5} + 1)i - 3\sqrt{5} - 7 & 10 & 6i + 8 & (\sqrt{5} + 1)i + 3\sqrt{5} - 7 \\
h_1h_2h_1^2h_2h_1h_2            & (-\sqrt{5} + 1)i - 3\sqrt{5} - 7 & -10 & -6i - 8 & (\sqrt{5} + 1)i + 3\sqrt{5} - 7 \\
\hline
h_1^2h_2h_1h_2                  & (2\sqrt{5} + 11)i + \sqrt{5} - 2 & 5i - 5 & -i + 7 & (-2\sqrt{5} + 11)i - \sqrt{5} - 2 \\
h_1h_2h_1h_2h_1^2h_2            & (3\sqrt{5} - 4)i + 4\sqrt{5} + 3 & -5i - 5 & 7i + 1 & (-3\sqrt{5} - 4)i - 4\sqrt{5} + 3 \\
h_2h_1h_2h_1^2h_2h_1h_2         & (\sqrt{5} - 4)i - 2\sqrt{5} + 3 & 5i - 15 & -7i - 1 & (-\sqrt{5} - 4)i + 2\sqrt{5} + 3 \\
h_2                             & (\sqrt{5} - 4)i - 2\sqrt{5} + 3 & 5i + 5 & 5i + 15 & (-\sqrt{5} - 4)i + 2\sqrt{5} + 3 \\
\hline
h_2h_1h_2h_1^2h_2h_1h_2h_1^2h_2 & (2\sqrt{5} + 11)i + \sqrt{5} - 2 & -5i + 5 & i - 7 & (-2\sqrt{5} + 11)i - \sqrt{5} - 2 \\
h_2h_1h_2h_1h_2                 & (3\sqrt{5} - 4)i + 4\sqrt{5} + 3 & 5i + 5 & -7i - 1 & (-3\sqrt{5} - 4)i - 4\sqrt{5} + 3 \\
h_1h_2h_1^2h_2                  & (\sqrt{5} - 4)i - 2\sqrt{5} + 3 & -5i + 15 & 7i + 1 & (-\sqrt{5} - 4)i + 2\sqrt{5} + 3 \\
h_2h_1^2h_2                     & (\sqrt{5} - 4)i - 2\sqrt{5} + 3 & -5i - 5 & -5i - 15 & (-\sqrt{5} - 4)i + 2\sqrt{5} + 3 \\
\hline
h_2h_1h_2                       & (-2\sqrt{5} + 11)i - \sqrt{5} - 2 & -5i - 5 & -7i - 1 & (2\sqrt{5} + 11)i + \sqrt{5} - 2 \\
h_2h_1^2h_2h_1h_2h_1^2h_2       & (-3\sqrt{5} - 4)i - 4\sqrt{5} + 3 & -5i + 5 & -i + 7 & (3\sqrt{5} - 4)i + 4\sqrt{5} + 3 \\
h_1^2h_2h_1h_2h_1^2h_2          & (-\sqrt{5} - 4)i + 2\sqrt{5} + 3 & -15i - 5 & i - 7 & (\sqrt{5} - 4)i - 2\sqrt{5} + 3 \\
h_1h_2                          & (-\sqrt{5} - 4)i + 2\sqrt{5} + 3 & 5i - 5 & -15i + 5 & (\sqrt{5} - 4)i - 2\sqrt{5} + 3 \\
\hline
(h_1h_2)^2                      & (-2\sqrt{5} + 11)i - \sqrt{5} - 2 & 5i + 5 & 7i + 1 & (2\sqrt{5} + 11)i + \sqrt{5} - 2 \\
h_1^2h_2                        & (-3\sqrt{5} - 4)i - 4\sqrt{5} + 3 & 5i - 5 & i - 7 & (3\sqrt{5} - 4)i + 4\sqrt{5} + 3 \\
(h_1h_2h_1^2h_2)^2              & (-\sqrt{5} - 4)i + 2\sqrt{5} + 3 & 15i + 5 & -i + 7 & (\sqrt{5} - 4)i - 2\sqrt{5} + 3 \\
h_2h_1h_2h_1^2h_2               & (-\sqrt{5} - 4)i + 2\sqrt{5} + 3 & -5i + 5 & 15i - 5 & (\sqrt{5} - 4)i - 2\sqrt{5} + 3 \\
\hline
\end{array}
\end{alignat*}
\caption{Coefficients of the $20$ locally equivalent states (scaled by a factor of $20$)
  which form a complete set of iso-entangled MUBs for two qubits. The
  first row corresponds to the fiducial vector given in Eq. (1). The ordering
  of the bases, separated by horizontal lines, is the same as in
  \eqref{eq:MUB1}--\eqref{eq:MUB5}. In the first column we list a
  group element in terms of the generators $h_1$ and $h_2$ that maps
  the first vector to the particular vector. Note that the first
  vector is an eigenvector of $h_1$.}
\label{table:isoMUB}
\end{table}

Furthermore, 
due to the symmetry of the group   $H_{\rm perm}$, for each tensor
factor the sets of $20$ unitary single-qubit matrices acting in both
subsystems to generate elements of all five MUBs from the  
fiducial state (1), form a unitary $5$-design. 
It is worth to emphasize here that a given configuration
treated as a design in various spaces  
 may lead to designs of a different degree.
For instance,  the set of five iso-entangled MUBs in ${\cal H}_4$
forms a projective $2$-design, 
the partial traces of these 20 vectors lead to a mixed-state $3$-design inside
 the Bloch ball  $\Omega_2$, while
the corresponding $20$ unitary matrices form an unitary $5$-design in $U(2)$.
A single iso-entangled basis is a projective $1$-design, 
its partial traces form a mixed-state $2$-design in $\Omega_2$,
and the corresponding 4 unitary matrices lead to a unitary $2$-design.


\section{Proof of Propositions}\label{proofprops}

\subsection{Proof of Proposition~\ref{prop1}}\label{APP_proof_1}

    Following the steps of the proof of an analogous statement for unitary designs  
    by Scott \cite{Sc08}, we start by introducing the following 
             operator in dimension $N^t$ determined by a constellation
             of $M$ states $\rho_i$ in dimension $N$:
                \begin{equation}\label{OpS}
                    S = \frac{1}{M}\sum_{i=1}^M \rho_i^{\otimes t} - \int_{\Omega_N} \rho^{\otimes t}\dd{\rho_{\rm HS}}.
                \end{equation}
           Next we consider the trace of the positive operator $S^\dag S$,
            \begin{align}
                0   & \leq \Tr(S^\dag S) = \nonumber\\ 
                    & = \frac{1}{M^2}\sum_{i,j=1}^M \Tr(\rho_i^{\otimes t}\rho_j^{\otimes t}) - 
                    2 \Tr(\frac{1}{M}\sum_{i=1}^M \rho_i^{\otimes t} 
                     \int_{\Omega_N} \sigma^{\otimes t} \dd{\sigma_{\rm HS}}) +
                       \Tr(\int_{\Omega_N} \rho^{\otimes t}\dd{\rho_{\rm HS}}
                         \int_{\Omega_N} \sigma^{\otimes t}\dd{\sigma_{\rm HS}}).
            \end{align}

            From this inequality we derive an analogue of the 
            Welch inequality for mixed-state $t$-designs:
            \begin{equation} \label{proofprop1}
                2 \Tr(\frac{1}{M}\sum_{i=1}^M \rho_i^{\otimes t} 
                \int_{\Omega_N} \sigma^{\otimes t}\dd{\sigma_{\rm HS}}) -
                \frac{1}{M^2}\sum_{i,j=1}^M \Tr(\rho_i^{\otimes t}\rho_j^{\otimes t})
                \leq
                \Tr(\int_{\Omega_N} \rho^{\otimes t}\dd{\rho_{\rm HS}}
                \int_{\Omega_N} \sigma^{\otimes t}\dd{\sigma_{\rm HS}}).
            \end{equation}
            Eq. \eqref{OpS} implies that  the above inequality is saturated
            if and only if the set of mixed states $\{\rho_i\}$ forms a mixed-state $t$-design, 
              which implies Proposition \ref{prop1} and leads to Eq. \eqref{welchprop1}. \qed 
              
              As a simple consequence of \eqref{proofprop1}, we can see that
			    every mixed-state $t$-design consisting of $M$ states satisfies
    			\begin{equation}
    			    \frac{1}{M^2}\sum_{i,j=1}^M \Tr(\rho_i\rho_j)^t = \gamma_{N,t},
    			\end{equation}
    			which is a necessary property of a mixed-state $t$-design.
           Here  $\gamma_{N,t} = \Tr \omega^2_{N,t}$
              denotes the purity of the  averaged state 
            $ \omega_{N,t}=
               \int_{\Omega_N} \rho^{\otimes t}\dd{\rho_{\rm HS}}$.

		\subsection{Purity of a random 2- and 3-quNit product state after twirling}
        We start with the case $t=2$ by evaluating a two-copy
        average product state.  
        It is convenient to use the twirling operation acting on any bipartite state $\rho_{AB}$ of dimension $N^2$, 
            defined by an integral with respect to 
            the Haar measure on the unitary group $U(N)$, corresponding to the local operations,
            \begin{equation}
                \mathcal{T}_2(\rho_{AB}) = \int_{U(N)} (U\otimes U)\rho_{AB}(U^\dag\otimes U^\dag) \dd_{\rm H}(U)  .
            \end{equation}
            The result of this operation can be given in terms of
            projection operators $\pi_\pm = (\mathbb{I}\pm
            O_{\text{SWAP}})/2$, projecting on symmetric and antisymmetric subspace respectively, as
            \begin{equation}
                \mathcal{T}_2(\rho_{AB}) = \alpha_+ \pi_+ + \alpha_- \pi_- ,
            \end{equation}
            with coefficients given by
            \begin{equation}
                \alpha_\pm = \frac{2\Tr(\rho_{AB}\pi_\pm)}{N(N\pm1)},
            \end{equation}
            with the SWAP-operation defined as $O_{\text{SWAP}}\qty(\ket{\psi}\otimes\ket{\phi}) = \ket{\phi}\otimes\ket{\psi}$.
        Making use of the fact that  $\ev{\sum_{i\neq j} \lambda_i\lambda_j} = 1 - \ev{\Tr\rho^2}$, 
         it is easy to show that 
            \begin{align}
                \ev{{\rm Tr}(\rho\pi_\pm)} = \frac{1 \pm \ev{\Tr\rho^2_{AB}}}{2},
            \end{align}
            where $\lambda_i$ with $i = 1,\hdots,N$ denote eigenvalues of $\rho$, while the average $\ev{\cdot}$ is taken over the entire set $\Omega_N$ of mixed states of size $N$ with respect to Hilbert-Schmidt measure.
            Let us consider $\rho_{AB}$ to be a
            diagonal bipartite product state, composed of two copies of a state
            in dimension $N$, $\rho_{AB} = \Lambda^{\otimes 2}$  with $\Lambda = \text{diag}(\lambda_1,\hdots,\lambda_N)$.  With this assumption, using the known
            average purity of the state $\rho$ \cite{ZS01} it can be shown that the
            coefficients $\alpha_\pm$ averaged with respect to the HS measure are then given by
            \begin{equation}
                \ev{\alpha_\pm} = \frac{N \pm 1}{N^3 + N}.
            \end{equation}
            Substituting $\alpha_\pm$ into the expression for
            $\mathcal{T}_2(\Lambda^{\otimes 2})$, we find the mean state $\omega_{2,N}$ given in terms of the twirled state averaged over the set $\Omega_N$
            \begin{equation}
                \omega_{N,2} = \ev{\mathcal{T}_2(\Lambda^{\otimes 2})} = \frac{N^2\mathbb{I} +  N O_{\text{SWAP}}}{N^4 + N^2} ,
                \label{omegaN2_gen}
            \end{equation}
            with purity given by 
            
            $$
            	\gamma_{N,2} = \frac{N^2+3}{\left(N^2+1\right)^2}.            
            $$   
\qed
            
            As an example, we give the simplest cases for $N=2,\,3$,
            which is a two qubit and two qutrit density matrix, respectively, symmetric with respect to the SWAP operation,
            \begin{equation*}
                \omega_{2,2} = \frac{1}{10}\mqty(3 & 0 & 0 & 0 \\
                                       			0 & 2 & 1 & 0 \\
                                        		0 & 1 & 2 & 0 \\
                                        		0 & 0 & 0 & 3),\qquad
                \omega_{3,2} = \frac{1}{30}\mqty(
						 4 & 0 & 0 & 0 & 0 & 0 & 0 & 0 & 0 \\
						 0 & 3 & 0 & 1 & 0 & 0 & 0 & 0 & 0 \\
						 0 & 0 & 3 & 0 & 0 & 0 & 1 & 0 & 0 \\
						 0 & 1 & 0 & 3 & 0 & 0 & 0 & 0 & 0 \\
						 0 & 0 & 0 & 0 & 4 & 0 & 0 & 0 & 0 \\
						 0 & 0 & 0 & 0 & 0 & 3 & 0 & 1 & 0 \\
						 0 & 0 & 1 & 0 & 0 & 0 & 3 & 0 & 0 \\
						 0 & 0 & 0 & 0 & 0 & 1 & 0 & 3 & 0 \\
						 0 & 0 & 0 & 0 & 0 & 0 & 0 & 0 & 4 \\
                ).
            \end{equation*}
            Purities of the states  $\omega_{2,2}$ and $\omega_{3,2}$ read $\gamma_{2,2} = 7/25$ and $\gamma_{3,2} = 3/25$.
       \medskip
 
  To obtain analogous results in the case of $t=3$ we need to deal
  with three-copy states and extend the set of permutation
  operators.  First step is to extend the twirling operation to three-partite systems by averaging local rotation $U$ over flat Haar measure of all three subspaces, such that it can be applied to any three-partite state $\rho_{ABC}$. By elementary consideration of symmetry it is found that a twirled tripartite state $\rho_{ABC}$ must be given by a linear combination of all permutation operators,
  such that
  the coefficients within conjugacy classes are the same,
    \begin{equation}
        \mathcal{T}_3(\rho_{ABC}) = \int_{U(N)} U^{\otimes 3}\rho_{ABC} U^{\dag\otimes 3}\dd_{\rm H}(U) = a_1 \mathbb{I} + a_2\qty(O_{(12)}+O_{(23)}+O_{(13)}) + a_3\qty(O_{(123)} + O_{(132)}),
        \label{t3ansatz}
    \end{equation}
   were $O_\sigma$ denotes the corresponding matrix representation of the permutation $\sigma$. 
   In particular, the twirling operation can be applied to three copies of the same local diagonal state $\Lambda$, and then averaged over all possible spectra. This is equivalent to averaging three-copy state $\rho^{\otimes3}$ over the entire space of mixed states $\Omega_N$ with respect to HS measure 
    \begin{equation}
        \omega_{N,3} = \ev{\mathcal{T}_3(\Lambda^{\otimes 3})} = \int_{\Delta_N}\int_{U(N)} U^{\otimes 3}\Lambda^{\otimes 3}U^{\dag\otimes 3} \dd{\Lambda_{\rm HS}}
        \dd_{\rm H}(U),
    \end{equation}
    which implies a system of three linear equations,
    \begin{align}
        1 = \Tr(\omega_{N,3})      & = N^3 a_1 + 3N^2 a_2 + 2N a_3 ,\nonumber\\
        \Tr(O_{(12)}\omega_{N,3})   & = N^2 a_1 + \qty(N^3 + 2N) a_2 + 2N^2 a_3, \nonumber \\
        \Tr(O_{(123)}\omega_{N,3})  & = N a_1 + 3N^2 a_2 + (N^3 + N a_3).
         \label{t3system}
    \end{align}

    First we consider the left-hand sides of the equations, given by well-known values \cite{ZS01},
    \begin{align}
        \Tr(O_{(123)}\omega_{N,3}) & = \ev{\Tr\rho^3} 
        = \frac{5N^2 + 1}{\qty(N^2+1)\qty(N^2 + 2)} .
    \end{align}
    
    In order to evaluate  $\Tr(O_{(12)}\omega_{N,3})$, 
    we will use properties of the permutation operator $O_{(12)}$, which imply
    \begin{align}
        \Tr(O_{(12)}\omega_{N,3}) = \ev{\Tr \rho^3} + \ev{\sum_{i\neq j}\lambda_i^2\lambda_j} = \ev{\sum_{i, j = 1}^N\lambda_i^2\lambda_j} = \ev{\sum_{i=1}^N \lambda_i^2 \sum_{j=1}^N \lambda_j} = \frac{2N}{N^2 + 1} .
    \end{align}
    
    Upon inserting these into equations into \eqref{t3system} we get
    \begin{align}
        a_1 & = \frac{N^3}{N^6 + 3N^4 + 2N^2}, &
        a_2 & = \frac{N^2}{N^6 + 3N^4 + 2N^2}, &
        a_3 & = \frac{N  }{N^6 + 3N^4 + 2N^2},
    \end{align}
    which  solves the case for $\omega_{N,3}$. In order to prove that a
    $3$-design is also a $2$-design we consider the partial trace over
    the third subsystem. It is obvious that $\Tr_3 \mathbb{I}_{N^3} =
    N \mathbb{I}_{N^2}$ and $\Tr_3 O_{(12)} = N O_{\text{SWAP}}$.  
    It is now easy to find that
    \begin{align}
        \Tr_3(O_{(13)}) & = \Tr_3(\sum_{i,j,k} \op{ijk}{kji}) = \sum_{i,j,k} \op{ij}{kj}\delta_{ik} = \mathbb{I}_{N^2},  \nonumber  \\
        \Tr_3(O_{(123)}) & = \Tr_3(\sum_{i,j,k} \op{ijk}{jki}) = \sum_{i,j,k} \op{ij}{jk}\delta_{ik} = O_{\text{SWAP}} .   \nonumber
    \end{align}
    Using this we obtain
    \begin{equation}
        \Tr_3(\omega_{N,3}) = \qty(Na_1 + 2a_2)\mathbb{I} + \qty(Na_2 + 2a_3)O_{\text{SWAP}} = \frac{N^2\mathbb{I} + NO_{\text{SWAP}}}{N^4+N^2},
    \end{equation}
    which is identical to \eqref{omegaN2_gen} and shows that a mixed-state
     design for $t=3$ is also $2$-design.
     By explicit calculation
    we obtain the desired coefficient  $\gamma_{N,3}$,
    \begin{equation}
        \gamma_{N,3} = \frac{N^6+9 N^4+24 N^2+2}{N \left(N^4+3 N^2+2\right)^2} .
    \end{equation}
             
\subsection{General scheme for calculating purity of averaged qNit states $\omega_{N,t}$} \label{APP_general_N}
 The approach for finding $\omega_{N,3}$ and $\gamma_{N,3}$ can be extended to any $t$. First we from similar observation that the twirled state of $t$ copies of diagonal local states $\mathcal{T}_t\qty(\Lambda^{\otimes t})$ is a sum over all permutation operators $O_\sigma$ with
 coefficients $a_i$, specific to conjugacy classes $C_i$:
    \begin{equation}
        \mathcal{T}_t\qty(\Lambda^{\otimes t}) = \int_{\Delta_N}\int_{U(N)} U^{\otimes t}\Lambda^{\otimes t}U^{\dag\otimes t}\dd{\Lambda_{\rm HS} \;\dd_{\rm H}(U) = \sum_{\qty{C_i}} a_i \sum_{\sigma\in C_i} O_\sigma}
    \end{equation}
 In order to compute the coefficients $a_i$ we need to solve the
 following system of linear equations obtained by considering an average twirled state $\ev{\mathcal{T}_t\qty(\Lambda^{\otimes t})}$
    \begin{align}
        1 &= \Tr(\omega_{N, t})  \nonumber \\
        \ev{\Tr(\rho^2)} & {}= \Tr(O_{(12)}\omega_{N, t})  \nonumber \\
        & \;\vdots  \nonumber \\
        \ev{\Tr(\rho^t)} & {}= \Tr(O_{(12\hdots t)}\omega_{N, t}) , \nonumber
    \end{align}
    where the left-hand sides can be obtained by similar arguments as for $t=3$.    
    We provide an Ansatz state that solves such system of equations for any given $t$
    \begin{equation}
        \omega_{N,t} = \frac{\sum_{\sigma_\in S_t} \Tr(O_\sigma) O_\sigma}{\sum_{\sigma_\in S_t}\Tr(O_\sigma)^2},
         \label{hypo2}
    \end{equation}
    and the expression for general $\gamma_{N,t}$ follows:
    \begin{equation}
        \gamma_{N,t} = \frac{\sum_{\sigma,\tau_\in S_t} \Tr(O_\sigma) \Tr(O_\tau) \Tr(O_{\sigma\tau})}{\sum_{\sigma,\tau_\in S_t}\Tr(O_\sigma)^2\Tr(O_\tau)^2}  .
        \label{general_gamma_eq}
    \end{equation}

Making use of formula \eqref{general_gamma_eq} one can derive further values of 
the coefficients $\gamma_{N,t}$, 
\begin{align}
    \gamma_{N,2} & = \frac{N^2+3}{\left(N^2+1\right)^2}, \\
    \gamma_{N,3} & = \frac{N^6+9 N^4+24 N^2+2}{N \left(N^4+3 N^2+2\right)^2}, \\
    \gamma_{N,4} & = \frac{N^8+18 N^6+123 N^4+344 N^2+90}{\left(N^6+6 N^4+11 N^2+6\right)^2}, \\
    \gamma_{N,5} & = \frac{N^{12}+30 N^{10}+375 N^8+2420 N^6+7422 N^4+3960 N^2+192}{N \left(N^8+10 N^6+35 N^4+50 N^2+24\right)^2} .
\end{align}

Due to relation (6), the above results allow to verify
whether a given constellation of density matrices forms a $t$-design.

\subsection{Proof of Proposition~\ref{prop2}}
\label{App_prop2}

It is known that the Fubini-Study measure in the space of pure states
in dimension $N^2$, related to the Haar measure on the group $U(N^2)$,
induces by partial trace the Hilbert-Schmidt measure on the reduced
space of mixed states \cite{ZS01}.

The density matrix corresponding to a pure state $\rho_\psi =
\ketbra{\psi}$ is linear in both the vector coordinates and their
conjugates.  Also its reduction $\rho_A = \Tr_B \ketbra{\psi}$ retains
this property. It is useful to think of the matrix $\rho_A$ as
decomposed in the canonical basis $\qty{\ket{i}\bra{j}}_{i,j=1}^N$ in the
space of matrices with some coefficients,
\begin{equation}
    \rho_A = \sum_{i,j=1}^N a^{ij}\ket{i}\bra{j}.
\end{equation}
The Schmidt decomposition of a bipartite state
\begin{equation}
    \ket{\psi} = \sum_{j=1}^{N} \sqrt{\lambda_j} \ket{j'}\otimes\ket{j''},
\end{equation}
which provides the eigenvalues $\lambda_i$ of the partial trace
$\rho_A$, may be viewed as a decomposition in a certain
basis. Therefore each eigenvalue $\lambda_i$ can be represented as
\nobreak
\begin{equation}
    \lambda_i = \sum_{k,l=1}^N A_i^{kl}a^{kl},
\end{equation}
where $A_i^{kl}$ is a transition matrix for the change of
basis. The above shows that every $\lambda_i$ is linear with respect
to the entries of the reduced matrix, which leads to conclusion that
it is linear with respect to the components of the pure state
$\ket{\psi}$.

Having established the proper class of polynomials $g_t$ of
eigenvalues of order $t$ and the flat Hilbert-Schmidt measure, we have
demonstrated that Proposition \ref{prop2} holds true.

\subsection{Reconstruction formula} \label{recoform}
In this section we demonstrate a way to obtain a reconstruction formula
for any state $\rho$ using measurements from a mixed $2$-design.  First,
in order to properly satisfy the requirements on tomography, we rescale
the design in such a way that
\begin{equation}
	\sum_{i=1}^M \tilde{\rho}_i = \mathbb{I}.
\end{equation}
which is satisfied by setting $\tilde{\rho}_i = \frac{N}{M}\rho_i$.
Given the requirement \eqref{omega} on mixed $2$-designs and result
\eqref{omegaN2_gen}, we arrive at the equation
\begin{equation}
    \frac{1}{M}\sum_{i=1}^M \rho_i^{\otimes 2} = \frac{M}{N^2}\sum_{i=1}^M \tilde{\rho}_i^{\otimes 2} = \frac{1}{N^4 + N^2}\sum_{j,k} N^2 \ketbra{j}\otimes\ketbra{k} + N\op{j}{k}\otimes\op{k}{j}
\end{equation}
Multiplying by an arbitrary operator $A\otimes\mathbb{I}$ and taking
the partial trace over the first system we obtain
\begin{equation}
    \frac{M}{N^2}\sum_{i=1}^M \Tr(A\tilde{\rho}_i)\tilde{\rho}_i = \frac{1}{N^4 + N^2}\qty(N^2 \Tr(A)\mathbb{I} + N A)
\end{equation}
Taking $A = \rho$ to be a density matrix, we easily get the reconstruction formula:
\begin{equation}
    \rho = \frac{\qty(N^2 + 1)M}{N}\sum_{i=1}^M p_i \tilde{\rho}_i - N \mathbb{I}_N.
\end{equation}
where $p_i = \Tr{\tilde{\rho}_i \rho}$. Note that a mixed-state design corresponds to a measurement of a 
bipartite system,
in which party $A$ does not have full control over the subsystem $B$.

\subsection{Proof of Proposition~\ref{dilution_prop}}
\label{app_dilut}
	By construction, averaging over two copies of each state in a projective 2-design yields a symmetric state
	\begin{equation}
		\sum_{i=1}^m \op{\psi_i}^{\otimes 2} = \frac{\mathbb{I}_{N^2} + O_{SWAP}}{N^2 + N} ,
	\end{equation}
	which, by elementary manipulation, is turned into a state corresponding to the defining state $\omega_{N, 2}$ for mixed 2-design
	\begin{equation}
		(1-p)\mathbb{I}_N^{\otimes 2} + p\sum_{i=1}^m \op{\psi_i}^{\otimes 2} = \frac{N^2\mathbb{I}_{N^2} + N O_{SWAP}}{N^4 + N^2} = \omega_{N, 2} ,
	\end{equation}
	which completes the proof.

\smallskip

In particular,  consider 
the standard  complete set of MUBs in the extended dimension $N^2$.
Then the  states obtained by reduction of the $N+1$ separable bases
form $N$ copies of the complete set of $N+1$ MUBs in ${\cal H}_N$.
Extending this configuration by the 
suitably weighted maximally mixed state, 
obtained by the partial trace of 
the remaining $N^2 - N$ maximally entangled basis,
one obtains  the mixed states 2-design  in dimension $N$ 
%
equivalent to the one implied by  Prop. \ref{dilution_prop}. 

\subsection{Proof of Proposition~\ref{prop4}}
\label{App_prop5}

Consider a simplicial $t$-design $\qty{\lambda_i}_{i=1}^n$ with
respect to the HS measure $\dd{\lambda}_{\rm HS}$ on the simplex of
eigenvalues $\Delta_N$, the corresponding set of diagonal matrices
$\Lambda_i = {\rm diag}\qty(\lambda_i)$ of order $N$, and a unitary
$t$-design $\qty{U_j}_{j=1}^m$ of matrices from $U(N)$.  By
definition, for any homogeneous function of order $t$ in the diagonal
entries of $\Lambda$ and the entries of $U$ and $U^\dag$, respectively,
evaluated and averaged over a design, is equal to the average over the
entire corresponding space,
\begin{align}
    \frac{1}{n} \sum_{i=1}^n f_t\qty(\lambda_i) & = \int_\Delta f_t(\lambda) \dd{\lambda}_{\rm HS}, &
    \frac{1}{m} \sum_{j=1}^m g_t\qty(U_j,U_j^\dag)
     &= \int_{U(N)} g_t\qty(U,U^\dag)  \dd_{\rm H}(U).
\end{align}

To construct a mixed-state $t$-design we will average a homogenous
function $h_t$ of degree $t$ over the space $\Omega_N$ of mixed states
with respect to the Hilbert-Schmidt measure $\mu_{\rm HS}$.  Such an
integral factorizes into the average over the space $U(N)$ of unitary
matrices with respect to the Haar measure $d_H$ and the average over
the simplex of eigenvalues $\Delta_N$ with respect to the HS measure
$\dd{\Lambda}_{\rm HS}$,
\begin{align}
    \int_{\Omega_N} h_t\qty(\rho)  \dd{\rho_{\rm HS}}
     = \int_{U(N)} \dd_{\rm H}(U) 
    \int_{\Delta_N}
     h_t\qty(U\Lambda U^\dag)  \dd{\Lambda}_{\rm HS}.
\end{align}

As the entries of a density matrix $U\Lambda U^\dag$ are linear in
the entries of $U$, $U^\dag$, and $\Lambda$, the function $h_t$ is
homogeneous of degree $t$ in the entries of $U$, $U^\dag$, and
$\Lambda$.  Hence the integral over the unitary group can be replaced
by the sum over the unitary design, while the remaining integral over the simplex is equal to the average
over the simplicial design
\begin{align}
    \int_{\Omega_N} h_t\qty(\rho) \dd{\rho_{\rm HS}}
    & = 
      \int_{\Delta_N}
     \sum_{j=1}^m h_t\qty(U_j\Lambda U_j^\dag) \dd{\Lambda}_{\rm HS}\nonumber \\
    & = \sum_{j=1}^m\sum_{i=1}^n h_t\qty(U_j\Lambda_i U_j^\dag) = \sum_{j=1}^m\sum_{i=1}^n h_t\qty(\rho_{ij}).
\end{align}
Thus, the expression for the mean value of
$h_t$ averaged over the entire set $\Omega_N$ implies that the set of
$mn$ density matrices obtained by a Cartesian product of both designs,
$\rho_{ij}=U_j \Lambda_i U_j^{\dagger}$, forms a mixed-state $t$
design.\hfill\qed

Note that the number $M=nm$ of elements of such a product design can
be reduced.  Let $\tilde \Delta_N$ denote the $1/N!$ part of the
simplex $\Delta_N$ distinguished by a given order of the components of
the probability vector, $\lambda_1 \ge \lambda_2 \ge \dots \ge
\lambda_N$.  Since unitary matrices contain permutations, which change
the order of the components $\lambda_i$, integrating over the spectrum
of $\rho$ it is possible to restrict the integration domain only to
the
set  $\tilde \Delta_N$.
Let $n'$ denotes the number of points of the simplicial design
belonging to the asymmetric part $\tilde \Delta_N$.
To obtain a mixed-state $t$-design $\Omega_N$ 
it is thus sufficient to consider the
 Cartesian product consisting of $n'm$ density matrices,
$\rho_{ij}=U_j\Lambda_i U_j^{\dagger}$, $i=1,\dots, n'$ and $j=1, \dots, m$.
If a vector $\lambda$  belongs to the boundary of  the chamber $\tilde \Delta_N$,
(see the example in Fig.~\ref{fig:d=3}),
one needs to weigh this point inversely proportional to the number of
chambers it belongs to.

Note also that the Platonic designs (see Supplemental Material \ref{platonic_des})
can be considered as a product of the HS $2$-design in 
$\tilde \Delta_2=[0,1/2]$,
consisting of a single point and shown in Fig.~3b,
and a unitary design in $U(2)$.
Due to the morphism between the groups $SU(2)$ and $SO(3)$,
the latter sets correspond to the
spherical designs on the sphere $S^2$,
which guarantee that the angular
distribution of the density matrices $\rho_i$
forming the mixed-state design is as uniform as possible.

The design corresponding to the  tetrahedral group
gives a tetrahedron inscribed inside the sphere of radius  $r=\sqrt{3/20}$,
which is unitarily equivalent any of constellations
obtained by partial trace of one of five iso-entangled bases
listed in table \ref{table:isoMUB}.
Thus the simplest mixed-state $2$-design consisting of four points
inside the Bloch ball is obtained by partial trace of 
one of the iso-entangled bases of size $d=N^2=4$.
It is thus natural to ask, whether this fact can be generalized
and  there exists a basis in the composite $N \times N$ system
such that the partial trace of the corresponding projectors 
forms a mixed-state $2$-design composed out of $N^2$ density 
matrices of size $N$.
Numerical results obtained for $N=3,4$ and $5$
suggest that this might be the case.

\section{Examples of mixed-state designs}

\subsection{Mixed-state designs in the Bloch ball}
\label{examples}

\begin{figure}[h]
    \centering
    \includegraphics[width=0.28\linewidth]{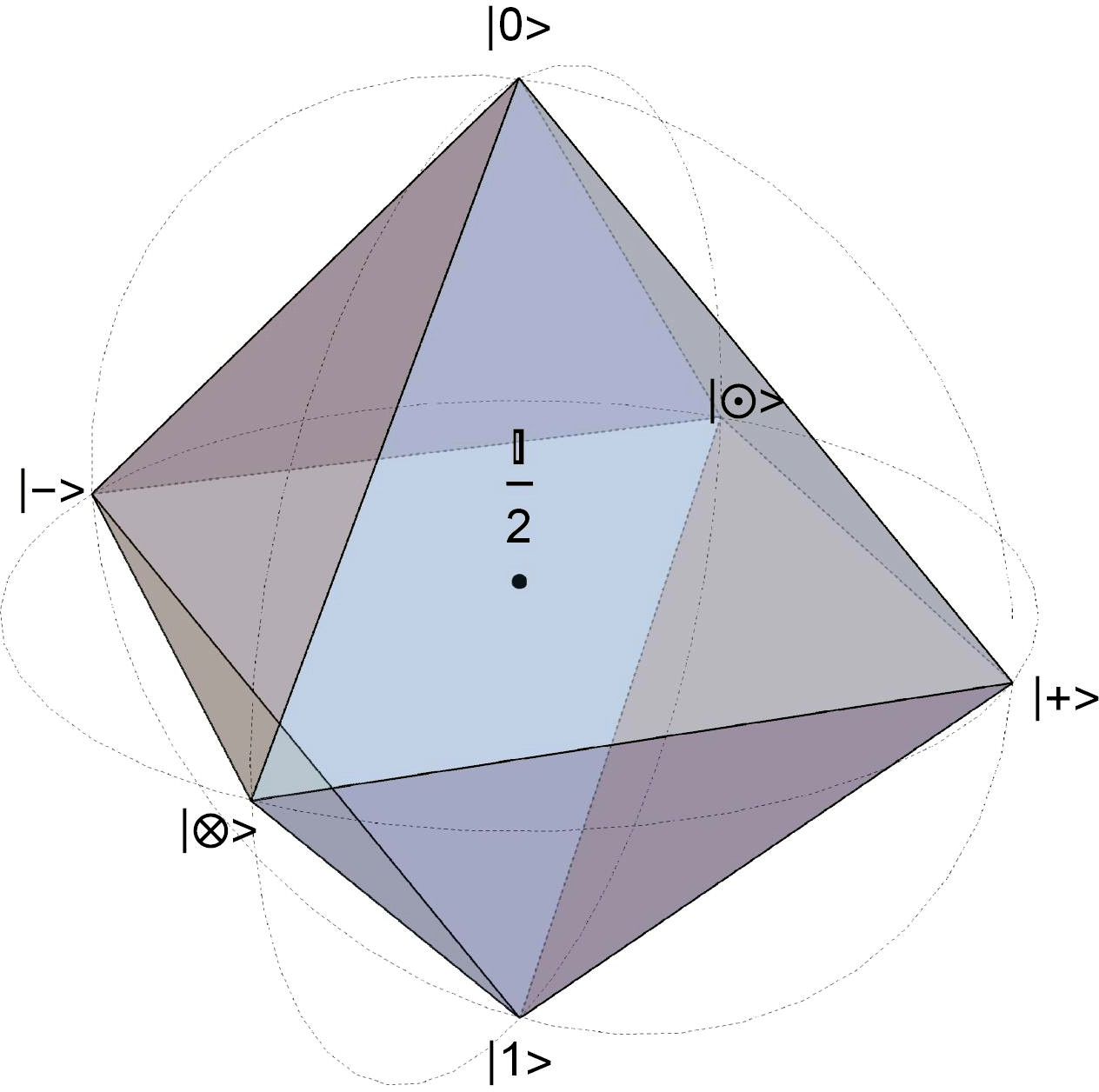}
    \caption{One-qubit mixed-state $2$-design obtained by partial
      trace of the standard set of $5$ MUBs in
      $\mathcal{H}_2^{\otimes2}$ consisting of $20$ points. Two
      maximally entangled bases induce a point of weight $8$ in the
      center of the Bloch ball, while each of the remaining three
      separable bases induces two antipodal points on the Bloch sphere
      with weight $2$ each.}
    \label{2210}
\end{figure}

\begin{figure}[h]
    \centering
    \includegraphics[width=0.6\linewidth]{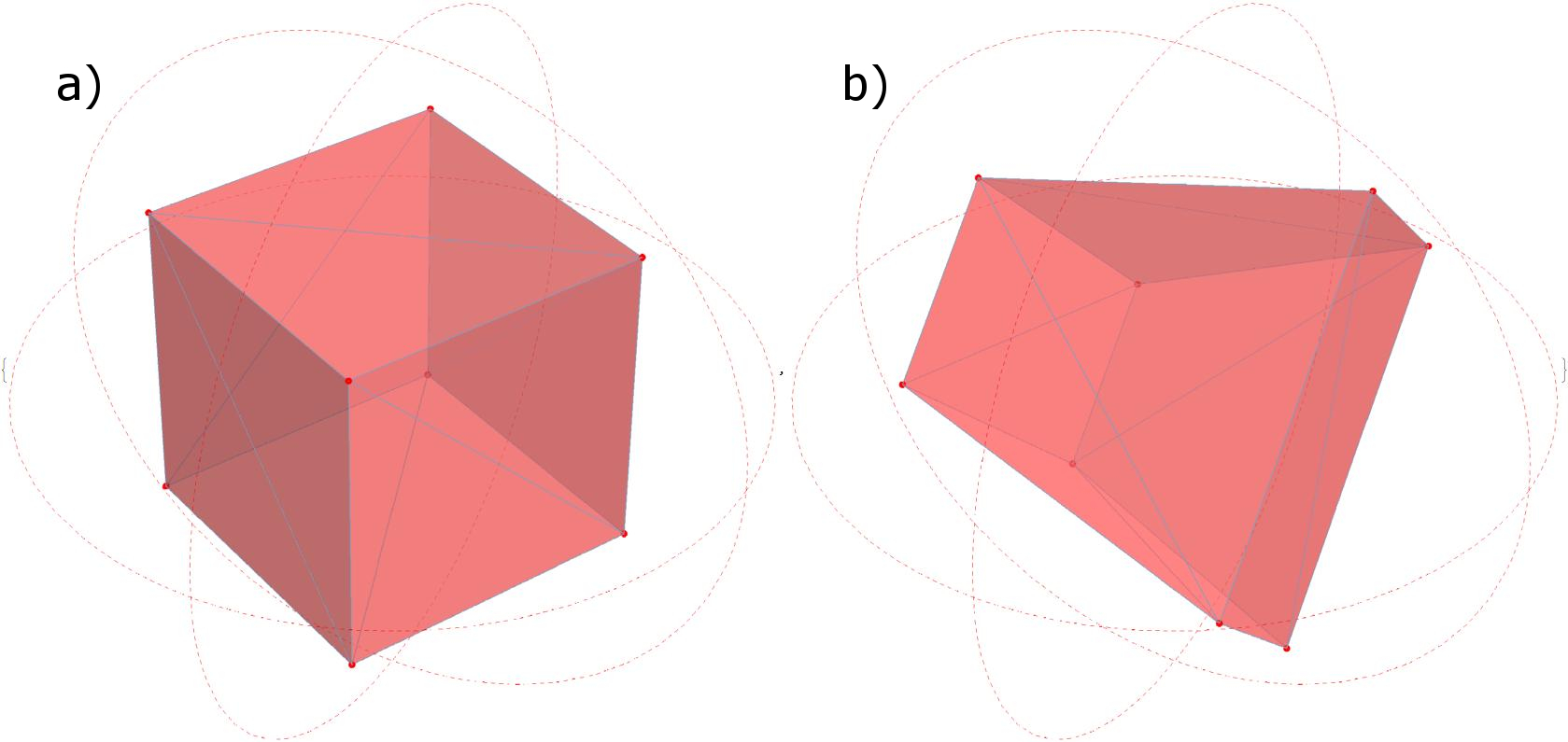}
    \caption{One-qubit mixed-state $2$-design composed of $8$ points,
      obtained by partial trace of $16$ pure states
      $\mathcal{H}_2^{\otimes 2}\ni\qty{\ket{\psi_i}}_{i=1}^{16}$
      forming the iso-entangled SIC-POVM for $2$ qubits
      \cite{ZTE10}. Both panels a) and b) present sets of $8$ points
      (each is doubly degenerated) belonging to the sphere of radius
      $\sqrt{3/20}$ inside the Bloch ball of radius $1/2$, which
      correspond to both partial traces.}
    \label{229}
\end{figure}

For the known mixed designs we compute the differences between the theoretical bound (6) and the ensemble value achieved, expressed as
        \begin{equation*}
            \delta_{N,t} = 
            \gamma_{N, t} 
               - 2 \Tr(\frac{1}{M}\sum_{i=1}^M \rho_i^{\otimes t} 
            \int_{\Omega_N} \sigma^{\otimes t}\dd{\sigma_{\rm HS}})
            +
            \frac{1}{M^2}\sum_{i,j=1}^M \abs{\Tr(\rho_i\rho_j)}^t ,
        \end{equation*}
        which are summarized in Table  \ref{tab:my_label}.

\begin{table}[h!]
            \centering
            \begin{tabular}{|c||c|c|c|c|c|c|c|}
            \hline
                t & 1 & 2 & 3 & 4 & 5 \\ \hline \hline
                Standard  MUB & 0 & 0 & 0 & 3.37$\times10^{-3}$ & 8.42$\times10^{-3}$ \\
                \textbf{IsoMUB} & 0 & 0 & 0 & 5.88$\times10^{-5}$ & 1.47$\times10^{-4}$ \\
                IsoSIC & 0 & 0 & 0 & 5.39$\times10^{-4}$ & 1.35$\times10^{-3}$ \\
                Witting Poly & 0 & 0 & 0 & 6.25$\times10^{-4}$ & 1.56$\times10^{-3}$  \\
                Hoggar Ex24 & 0 & 0 & 0 & 3.37$\times10^{-3}$ & 8.42$\times10^{-3}$  \\
                \hline
            \end{tabular}
            \caption{Values of the differencies $\delta_{2,t}$ with respect to the global
              extremum $ \gamma_{N, t}$ computed
            for known one-qubit mixed designs. Since condition (6) is satisfied,
              $\delta_{2,t}=0$ for $t=1,2,3$, all
             these constellations of density matrices form mixed-state
              $3$-designs. Different values for $t = 4,5$ are implied
             by differences between the designs. It is easily seen
             that IsoMUB is the closest solution to be  a
             $4$-design. Moreover, one may observe that values for
             the standard MUB and  the example 24  of Hogar \cite{H82} are identical, as
             implied by equivalence of the induced mixed designs.}
            \label{tab:my_label}
\end{table}

\begin{figure}[h]
    \centering
    \includegraphics[width=0.28\linewidth]{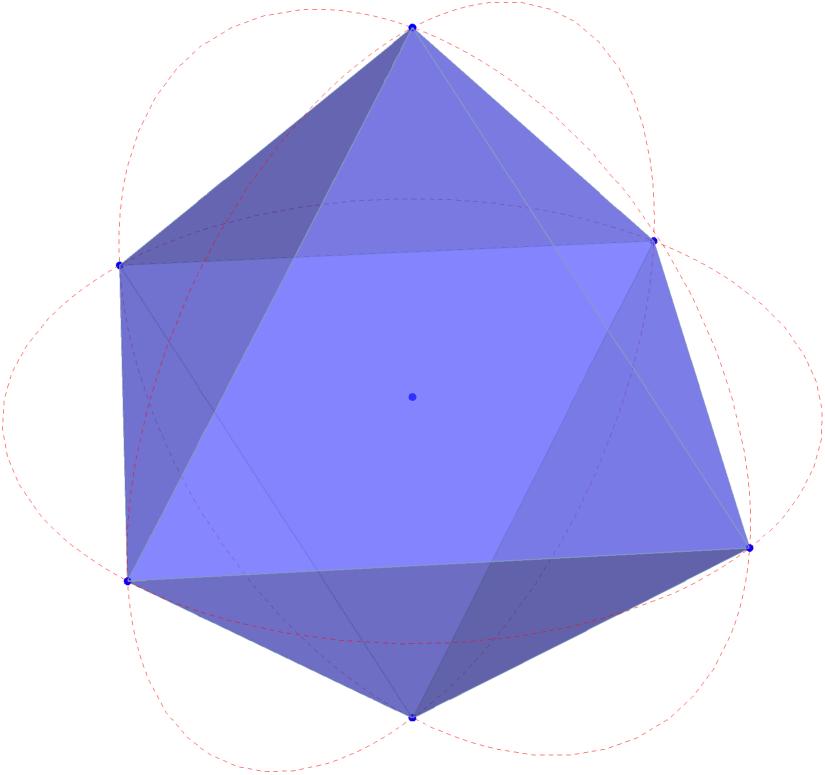}
    \caption{One-qubit mixed-state $3$-design obtained by partial
      trace of the projective $3$-design consisting of $60$ states in
      $\mathcal{H}_4$ provided by Hoggar \cite{H82}
      in his Example 24. 
    }
    \label{2310}
\end{figure}

\begin{figure}[h]
    \centering
    \includegraphics[width=0.49\linewidth]{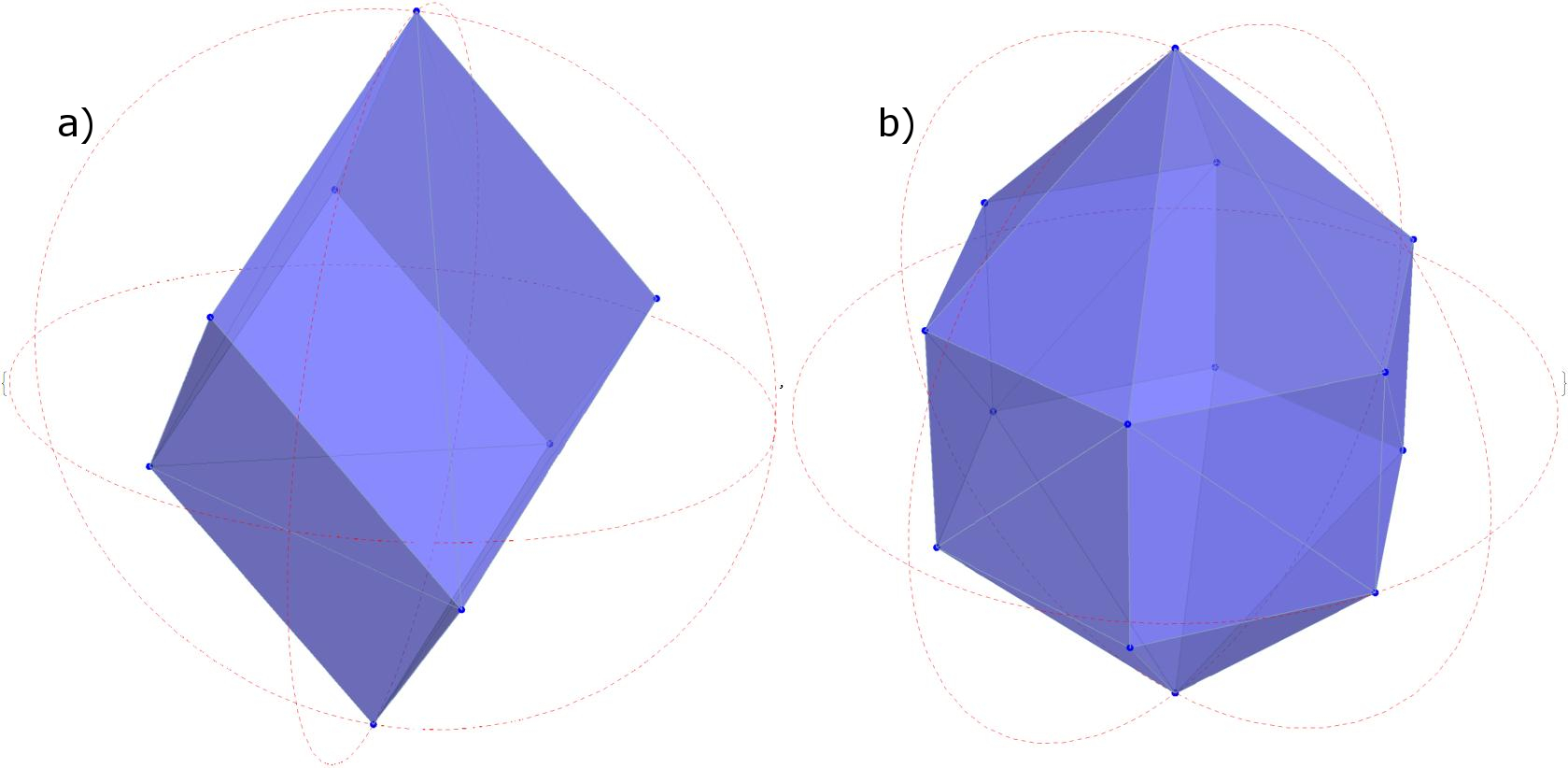}
    \caption{One-qubit mixed-state $3$-design obtained by partial
      trace of $40$ states in $\mathcal{H}_4$ leading to the Witting
      polytope, consisting of $8$ points in reduction A and $14$ in
      reduction B. 
    The points at the poles in reduction A have weight $1$ and the remaining $6$ points have weight $3$ each.
    In reduction B, points at the poles have weight $2$ while the remaining $12$ points have weight $3$ each.}
    \label{2320}
\end{figure}

In the case of the Witting polytope (which is equivalent to the
Penrose dodecahedron \cite{WA17}) we have two regular figures---a
parallelepiped (a) and an elongated bipyramid (b) in respective
reductions. This suggests that properly resized regular polytopes
could serve as templates for $t$-designs of different orders.

\medskip

\subsection{Platonic designs}
\label{platonic_des}
In this section we consider constellations of states derived from Platonic solids and their relation with mixed-state $t$-designs.

One may consider sets of states in $\mathcal{H}_2$ derived from any
$3$-dimensional solid via the standard form of a pure state of a qubit
\begin{equation}
    \ket{\psi(\theta,\phi)} = \mqty(\cos\frac{\theta}{2} \\ e^{i\phi}\sin\frac{\theta}{2}).
\end{equation}
Using its antipodal counterpart
\begin{equation}
    \ket{\tilde{\psi}(\theta,\phi)} = \mqty(\sin\frac{\theta}{2} \\ -e^{i\phi}\cos\frac{\theta}{2})
\end{equation}
one can interpolate between the maximally mixed state and the pure state:
\begin{equation}
    \rho(\theta,\phi,a) = a\ketbra{\psi} + (1-a)\ketbra{\tilde{\psi}}.
\end{equation}
Using this, we have found that for each Platonic solid there exists a
corresponding mixed-state $2$-design, given by $a = \frac{1}{10}
\left(5-\sqrt{15}\right)$. The analytic form of the tetrahedral
design $\qty{\rho_i}_{i=1}^4$, corresponding to a rescaled SIC-POVM, is given below:

\begin{align*}
    \rho_1 & = \left(
	\begin{array}{cc}
	\frac{1}{10} \left(5-\sqrt{15}\right) & 0 \\
	0 & \frac{1}{10} \left(5+\sqrt{15}\right) \\
	\end{array}
	\right), &
	\rho_2 & = \left(
	\begin{array}{cc}
		\frac{1}{30} \left(15+\sqrt{15}\right) & e^{-i \frac{2\pi}{3}} 		\sqrt{\frac{2}{15}} \\
	e^{i \frac{2\pi}{3}} \sqrt{\frac{2}{15}} & \frac{1}{30}\qty(15 - \sqrt{15}) \\
	\end{array}
	\right),
	\\
	\rho_3 & = \left(
	\begin{array}{cc}
	 \frac{1}{30} \left(15+\sqrt{15}\right) & -\sqrt{\frac{2}{15}} \\
	 -\sqrt{\frac{2}{15}} & \frac{1}{30}\qty(15 - \sqrt{15}) \\
	\end{array}
	\right), &
	\rho_4 & =\left(
	\begin{array}{cc}
	  \frac{1}{30} \left(15+\sqrt{15}\right) & e^{i \frac{2\pi}{3}} \sqrt{\frac{2}{15}} \\
	 e^{-i \frac{2\pi}{3}} \sqrt{\frac{2}{15}} & \frac{1}{30}\qty(15 - \sqrt{15}) \\
	\end{array}
	\right).
\end{align*}

As mentioned in the main body of the paper,
this configuration of four mixed states 
is equivalent up to a unitary rotation 
to the $2$-designs
obtained by partial trace of any of five iso-entangled bases
given in Table \ref{table:isoMUB} 
and shown in Fig. \ref{2221}.

\begin{table}[h!]
            \centering
            \begin{tabular}{|c|c|c|c|c|c|c|c|}
            \hline
                t & 2 & 3 & 4 & 5 \\ \hline \hline
                Tetrahedral & 0 & 6$\times10^{-3}$ & 1.25$\times10^{-2}$ & 1.69$\times10^{-2}$ \\
                Octahedral & 0 & 0 & 1.14$\times10^{-3}$ & 2.85$\times10^{-3}$ \\
                Cubic (IsoSIC) & 0 & 0 & 5.39 $\times10^{-4}$ & 1.35$\times10^{-3}$ \\
                Icosahedral & 0 & 0 & 5.88$\times10^{-5}$ & 1.47$\times10^{-4}$ \\
                Dodecahedral (IsoMUB) & 0 & 0 & 5.88$\times10^{-5}$ & 1.47$\times10^{-4}$ \\
                \hline
            \end{tabular}
            \caption{Residual values of $\delta_{2,t}$   for $t\geq 2$ for mixed $2$-designs derived from Platonic solids.
 By construction  the value $\delta_{2,t}=0$ implies that the constellation forms a $t$-design
 in the Bloch ball $\Omega_2$.
  Note that the icosahedral and dodecahedral configurations form $3$-designs,
  while their residual values  $\delta_{2,4}$  and $\delta_{2,5}$ are identical.}
            \label{tab:my_label2}
\end{table}

\section{Projective designs and averaging sets in the probability simplex}\label{APP_interval}
In this section we shall construct averaging sets on the
$(N-1)$-dimensional simplex $\Delta_N$ containing all probability
vectors of size $N$.  In the simplest case of $N=2$ we consider
collections of points from the interval $[-1/2,1/2]$. Such designs
with respect to the flat Lebesgue measure are related to projective
designs for $N=2$, while those with respect to the Hilbert-Schmidt
measure allow one to find the radius of the sphere inside the Bloch
ball, at which points forming a symmetric mixed-state design should
be located.

\begin{figure}[h]
    \centering
    \includegraphics[width = .67\linewidth]{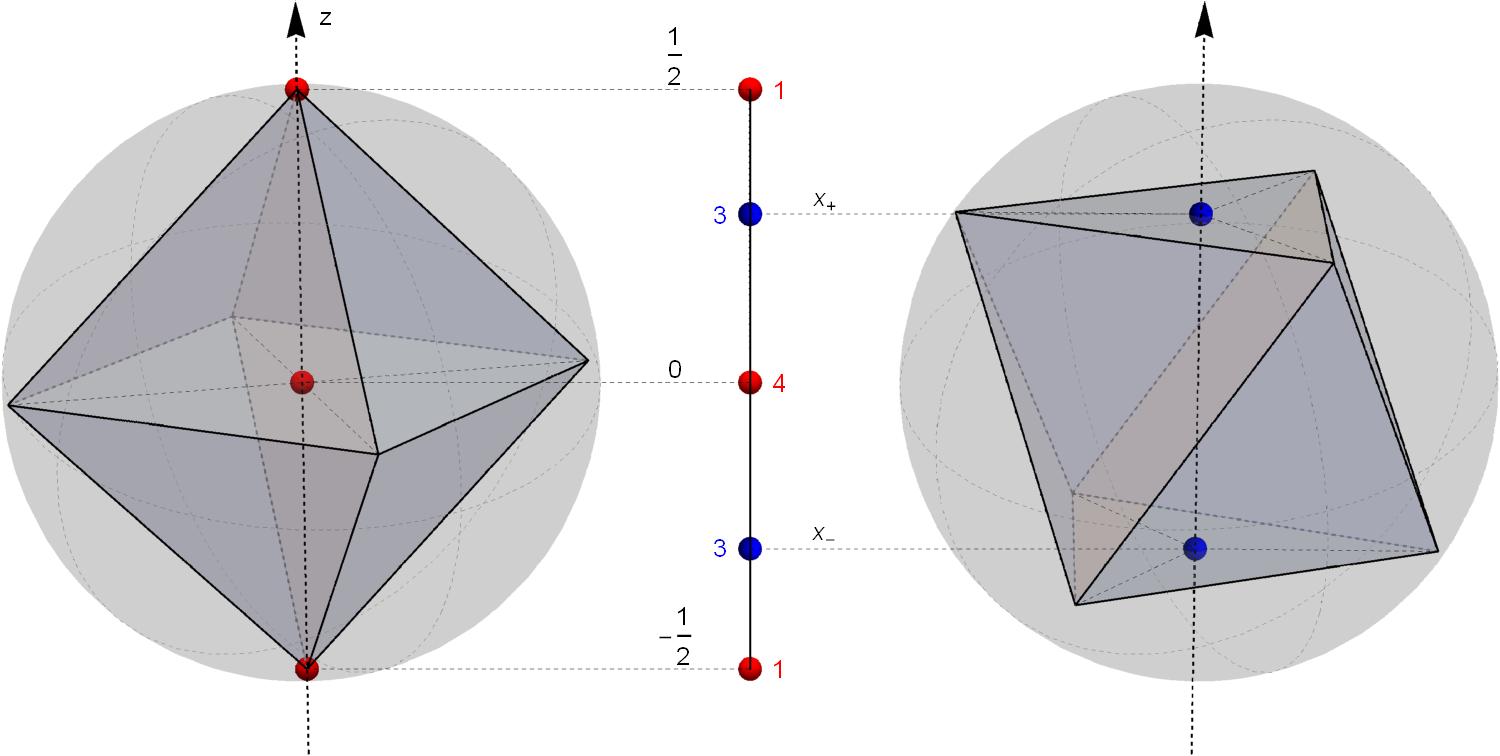}
    \caption{Averaging sets in the interval $[-1/2,1/2]$
   obtained by decoherence of the particular  constellations of  one-qubit pure states:
    a) regular octahedron representing  three MUBs in dimension $N=2$ leads 
    to three red points with weights $1,4,1$ corresponding to 
    Simpson integration rule; b) different projection of the same
     projective $2$-design leads to the design consisting of two blue points  $x_{\pm} = \pm \   {\sqrt{3}}/6$,
       which corresponds to the Gauss-Legendre integration rule.
}
    \label{new_Jakub}
\end{figure}

A link between projective designs consisting of pure
states of an $N \times N$ system and 
designs formed by the set of density matrices of size $N$
was established in Proposition 2.
This result can be treated as an example of a more general construction:
an averaging set for a certain space $\Xi$ with respect to the measure $\mu$
allows one to find a corresponding design on the space $\Omega =T(\Xi)$ 
with respect to the image measure ${\mu}'$ (also called push-forward measure)
 induced by the transformation $T$. 
 More precisely, for any measurable set $K \subset \Omega$ 
 its image measure reads  $\mu'(K)=\mu(T^{-1}(K))$.
 In the case considered here 
$\Xi$ represents the complex projective space ${\mathbbm C}P^{\;N^2-1}$,
while  $T$ denotes the partial trace over an $N$ dimensional subsystem,
and $\Omega$ represents the set of density matrices of size $N$.

In a similar way one can consider spherical designs on the Bloch
sphere, $S^2={\mathbbm C}P^{\;1}$ and analyze their projections onto an
interval, Fig.~3.  Further examples of averaging sets
on the interval induced by spherical designs are shown in
Fig. \ref{new_Jakub}.  A regular octahedron inscribed in a sphere with
two vertices at the antipodal poles and four on the equator (see
panel a)) induces by projection an averaging set on the interval
with weights $1,4,1$ and corresponds to the Simpson rule for numerical
integration.  Another projection of the octahedron on a line leads to
a set consisting of two points at $x_{\pm}=\pm 1/2\sqrt{3}$
corresponding to the $2$-point Gauss-Legendre integration rule in 
$[-1/2,1/2]$, see
Fig. \ref{new_Jakub}b.

In physical terms such a projection of the Bloch sphere
onto a line describes decoherence due to interaction of the
system with environment. It is then fair to say that 
any one-qubit projective design decoheres
to a design  on an interval,
while a projective design formed by pure states in dimension $d$
is mapped by the coarse-graining map (dephasing channel),
$T\colon \ket{\psi}\bra{\psi} \mapsto  {\rm diag} (\ket{\psi}\bra{\psi})$,
to an averaging set  in the simplex of $d$-point probability vectors.
Such a configuration forms a design in the simplex 
with respect to the flat Lebesgue measure,
which is an image of the unitarily invariant Fubini-Study measure 
on the complex projective space ${\mathbbm C}P^{\;d-1}$
with respect to the  coarse-graining map \cite{ZS01}.

For completeness, we present here the explicit form of $t$-designs on
the interval for some low values of $t$.  Working out conditions
(10) for the Lebesgue measure on $[-1/2,1/2]$ it is easy to
check whether a set consisting of $M$ points leads to a $t$-design.
In some cases one may even get more than required: the set of $M=2$
points satisfies not only the condition for a $2$-design, but also for
a $3$-design.
    \begin{align}
        t=1,\,M=1:\quad x_1 & = 0, \\
        t=3,\,M=2:\quad x_1 & = -  \frac{1}{2\sqrt{3}}, & 
        x_2 & = \frac{1}{2\sqrt{3}}, \\
        t=3,\,M=3:\quad x_1 & = - \frac{1}{2\sqrt{2}}, &
        x_2 & = 0, &
        x_3 & = \frac{1}{2\sqrt{2}}, \\
        t=5,\,M=4:\quad x_1 & = -\frac{1}{30} \sqrt{75+30 \sqrt{5}}, &
        x_2 & = -\frac{1}{30} \sqrt{75-30 \sqrt{5}}, \nonumber \\
        x_3 & = \frac{1}{30} \sqrt{75-30 \sqrt{5}}, &
        x_4 & = \frac{1}{30} \sqrt{75+30 \sqrt{5}}, \\
        t=5,\,M=5:\quad x_1 & = - \frac{1}{12}\sqrt{15 + 3\sqrt{11}}, &
        x_2 & = - \frac{1}{12}\sqrt{15-3 \sqrt{11}}, &
        x_3 & = 0, \nonumber \\
        x_4 & = \frac{1}{12}\sqrt{15-3 \sqrt{11}}, &
        x_5 & = \frac{1}{12}\sqrt{15 + 3\sqrt{11}}.
    \end{align}
    
    Averaging sets on an interval with respect to the Hilbert-Schmidt measure
    are related to mixed-state designs in the set of one-qubit density matrices.
    In particular, $2$-design corresponds to the projection of the
    cube inscribed into the sphere of radius  $r=\sqrt{3/20}$ located
    inside the Bloch ball, see Fig.~3d.  
    
    \begin{align}
        t=3,\,M=2:\quad \lambda_1 & = -\sqrt{\frac{3}{20}}, & \lambda_2 & = 
        \sqrt{\frac{3}{20}}. \label{2points} \\
        t=3,\,M=3:\quad \lambda_1 & = -\frac{3}{2 \sqrt{10}}, & \lambda_2 & = 0, & \lambda_3 & = \frac{3}{2 \sqrt{10}}. \label{3points} \\
        t=5,\,M=4:\quad \lambda_1 & = -\frac{1}{70} \sqrt{735+70 \sqrt{21}}, & \lambda_2 & = -\frac{1}{70} \sqrt{735-70 \sqrt{21}}, \\
        \lambda_3 & = \frac{1}{70} \sqrt{735 -70 \sqrt{21}}, & \lambda_4 & = \frac{1}{70} \sqrt{735 + 70 \sqrt{21}}. \label{4points}
    \end{align}

Note that the above results can be used to search for one-qubit
mixed-state $t$-designs with $t\ge 3$ as $r_i=|1/2-x_i|$ denotes radii
of the spheres inscribed inside the Bloch ball containing density
matrices belonging to the design.

\subsection{Quantum states and designs in the triangle of $3$-point probability distributions}
\label{d3}

The standard solution for a complete set of MUB in dimension $d=3
\times 3$ system consists of $4$ separable bases and $6$ maximally
entangled bases \cite{La04}.  The partial trace of these $10\times 9$ pure
states of size $9$ leads to a collection of $90$ density matrices of
size $3$, which due to Proposition~2 generates a mixed-state
$2$-design in the set $\Omega_3$.  Eigenvalues of these states form a
$2$-design in the probability simplex with respect to the
Hilbert-Schmidt measure, induced by partial trace, see
Fig.~\ref{fig:d=3}a.
Note that these $3$-point probability
distributions represent Schmidt vectors of the original pure states
$|\Psi_j\rangle$ of the bipartite system composed of two qutrits.

\begin{figure}[h]
    \centering
    \includegraphics[width = .33\linewidth]{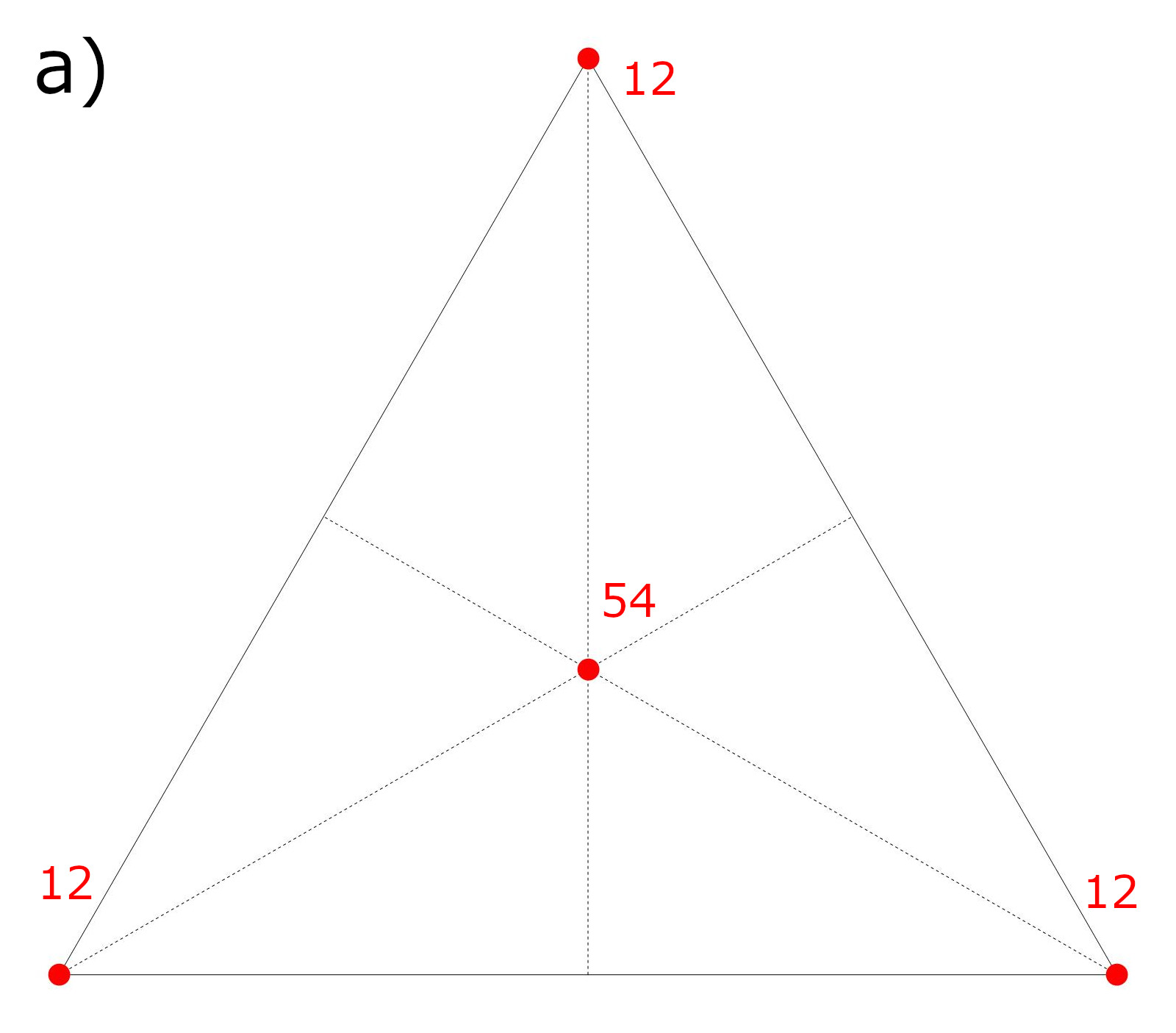}
    \includegraphics[width = .33\linewidth]{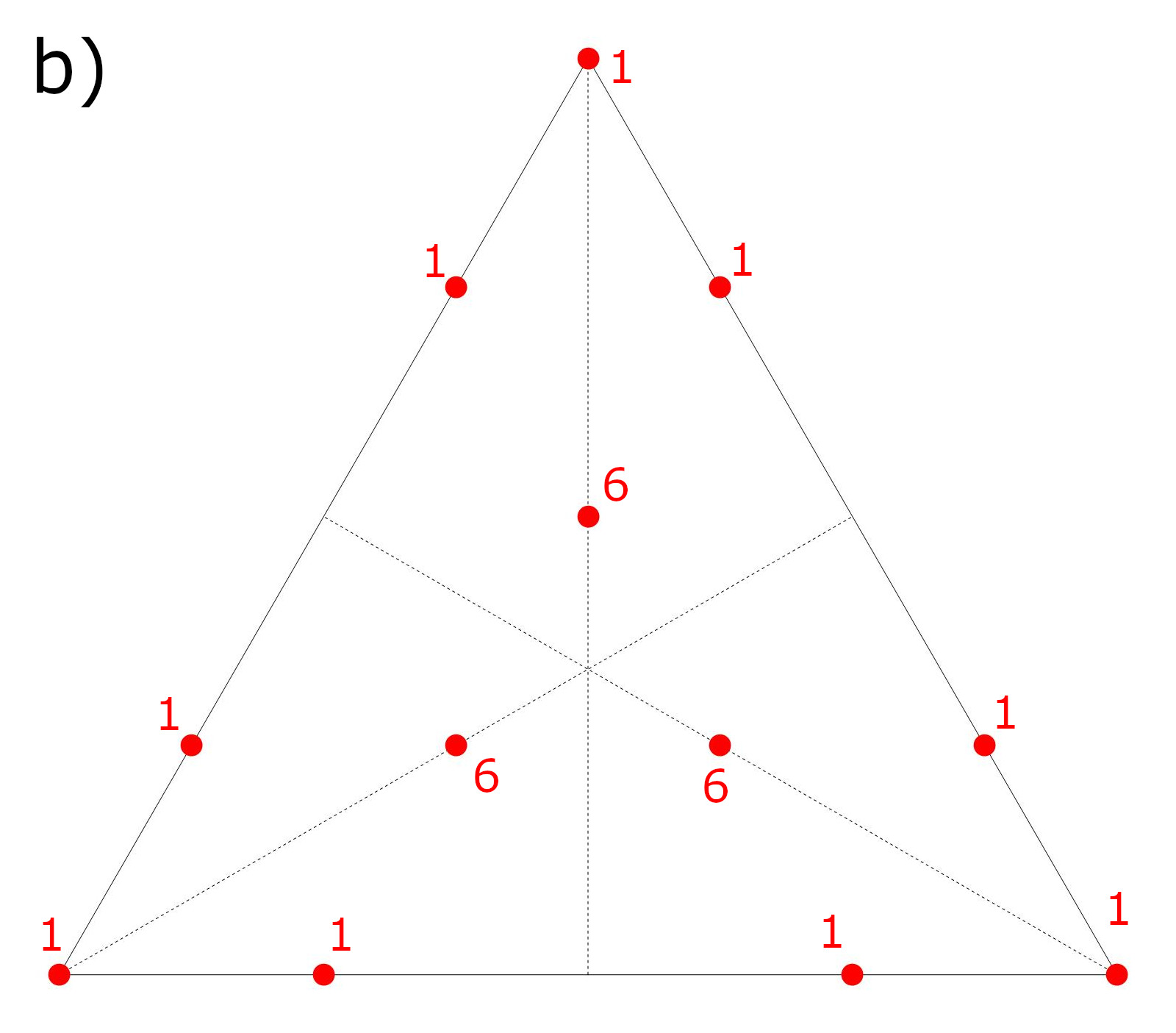}
    \caption{Designs in the simplex
    of $N=3$ probability distributions: a) HS $2$-design
    obtained by reduction of $90$ states forming the standard set of $10$ MUBs in $\mathcal{H}_9$ \cite{La04}. 
    Points at the vertices, corresponding to pure states, are obtained
    from $4$ separable bases, while the remaining $6$ maximally
    entangled bases yield the center point representing the
    maximally mixed state with spectrum $(1/3,1/3,1/3)$. The numbers
    in red denote weights assigned to each point, adding up to the
    total $90$, the total number of states. 
  b) $2$-design with respect to the Lebesgue measure obtained
  by taking diagonal elements of projections onto $9$ states forming a
  SIC in dimension $N=3$. 
}
    \label{fig:d=3}
\end{figure}

To obtain a $2$-design in this probability simplex with respect to the
flat measure it is sufficient to take an arbitrary realization of a
projective $2$-design in the set $\Xi_3$ of pure states in dimension
$d=3$ and take the corresponding classical states.
Figure~\ref{fig:d=3}b shows such a configuration in the simplex, which
stems from $9$ states forming a SIC in dimension three.

In the similar spirit, the coarse-graining map, corresponding to
complete decoherence and sending projectors onto pure states to
classical probability vectors, applied to any SIC configuration in
dimension $d=4$ produces a $2$-design with respect to the Lebesgue
measure in the regular tetrahedron of $4$-point probability
distributions.  On the other hand, the Schmidt vectors of $64$ pure
states forming a SIC for two subsystems with four levels each lead to
a $2$-design with respect to the Hilbert-Schmidt measure induced by
partial trace \cite{ZS01}.

\end{widetext}


\begin{thebibliography}{99}



\bibitem{AJK04} J. B. Altepeter,  D. F. V. James, and P. G. Kwiat, 
Quantum state tomography,
 in M. Paris, J. Reha{\v c}ek (eds.) {\sl Quantum state estimation},
(Springer-Verlag; Berlin, 2004). 

\bibitem{BK10} R. Blume-Kohout, 
Optimal, reliable estimation of quantum states,
{\sl New J. Phys.} {\bf 12}, 043034 (2010).

 \bibitem{RBSC04} J. M. Renes, R. Blume-Kohout, A. J. Scott, and
   C. M. Caves, Symmetric informationally complete quantum
   measurements, J. Math. Phys. {\bf 45}, 2171  (2004).

\bibitem{S06} A. J. Scott, Tight informationally complete quantum measurements, 
J. Phys. A {\bf 39}, 13507 (2006).
  
\bibitem{WF89} W. Wootters and B. Fields, 
Optimal State-Determination by Mutually Unbiased Measurements,
Ann. Phys. {\bf 191}, 363 (1989).
   

\bibitem{AS10} B. A. Adamson and A. M. Steinberg,
Improving Quantum State Estimation with Mutually Unbiased Bases,
{\sl Phys. Rev. Lett.} {\bf  105}, 030406 (2010).

\bibitem{SG10} A. J. Scott and M. Grassl, SIC-POVMs: A new computer study,
{\sl  J.~Math. Phys.} {\bf  51}, 042203 (2010).

\bibitem{GS17} M. Grassl and A. J. Scott, 
Fibonacci-Lucas SIC-POVMs, 
{\sl J. Math. Phys.} {\bf 58}, 122201 (2017).

\bibitem{BBELTZ07} I. Bengtsson, W. Bruzda, A. Ericsson,
J.-{\AA}. Larsson, W. Tadej, and K. {\.Z}yczkowski,
MUBs and Hadamards of Order Six, 
{\sl J. Math. Phys.} {\bf 48}, 052106  (2007).


 \bibitem{DEBZ10} T. Durt, B.-G. Englert, I. Bengtsson, and K. {\.Z}yczkowski,
 On mutually unbiased bases, 
{\sl Int. J. Quantum Information} {\bf 8}, 535 (2010).

\bibitem{L78} E. Lubkin, Entropy of an $n$‐system from its correlation
  with a $k$‐reservoir, J. Math Phys. {\bf 19} 1028 (1978). 

\bibitem{ZTE10} H. Zhu, Y. S. Teo, and B.-G. Englert,
 Structure of Two-qubit Symmetric Informationally Complete POVMs,
 {\sl Phys. Rev. A} {\bf 82}, 042308 (2010).

  \bibitem{LLZZ18}  Z-W. Liu, S. Lloyd, E. Y. Zhu, and H. Zhu,  
  Generalized Entanglement Entropies of Quantum Designs,
{\sl Phys. Rev. Lett.} {\bf 120}, 130502 (2018).

\bibitem{CGZ18} J. Czartowski, D. Goyeneche and K. {\.Z}yczkowski,
Entanglement properties of multipartite informationally complete quantum  measurements,
{\sl J. Phys. A} {\bf 51}, 305302 (2018).

\bibitem{La04} J. Lawrence, Mutually unbiased bases and trinary operator sets for $N$ qutrits,
{\sl Phys. Rev. A} {\bf 70}, 012302 (2004).

\bibitem{RBKS05} J. L. Romero, G. Bj{o}rk, A. B. Klimov, and L. L. S{\'a}nchez-Soto,
Structure of the sets of mutually unbiased bases for $N$ qubits,
{\sl Phys. Rev. A} {\bf 72}, 062310 (2005).

\bibitem{WPZ11} M. Wie{\'s}niak,  T. Paterek, and A. Zeilinger,
Entanglement in mutually unbiased bases,
{\sl New J. Phys.} {\bf 13}, 053047 (2011).

\bibitem{GG15} D. Goyeneche, S. Gomez, 
Mutually unbiased bases with free parameters,  Phys. Rev. A 92, 062325 (2015).

 \bibitem{Re05} J. Renes, 
Equiangular spherical codes in quantum cryptography,
{\sl Quantum Inf. Comput.} {\bf 5}, 81 (2005).

\bibitem{Ta12} G. N. M. Tabia, Experimental scheme for qubit and
  qutrit symmetric informationally complete positive operator-valued
  measurements using multiport devices, 
{\sl Phys. Rev. A} {\bf 86}, 062107 (2012).

\bibitem{DGS77} P. Delsarte, J. Goethals, and J. Seidel, 
Spherical codes and designs,
{\sl  Geom. Dedicata} {\bf  6},  363 (1977).

 \bibitem{BZ17} I. Bengtsson and K. {\.Z}yczkowski,
{\sl Geometry of Quantum States}, II edition,
 (Cambridge University Press, 2017).

\bibitem{GAE07} D. Gross, K. Audenaert, and J. Eisert, 
Evenly distributed unitaries: on the structure of unitary designs,
{\sl  J. Math. Phys.} {\bf  48}, 052104 (2007).

\bibitem{Sc08} A. Scott, 
Optimizing quantum process tomography with unitary $2$-designs,
{\sl  J. Phys. A} {\bf 41}, 055308 (2008).

 \bibitem{RS09} A. Roy and A. J. Scott, 
Unitary designs and codes,  
{\sl Des. Codes Cryptogr.} {\bf  53}, 13 (2009).

\bibitem{DCEL09} C. Dankert, R. Cleve, J. Emerson, and E. Livine,
Exact and Approximate Unitary $2$-Designs: Constructions and Applications,
{\sl Phys. Rev. A} {\bf 80}, 012304 (2009).

 \bibitem{GKK15}  D. Gross, F. Krahmer, and R. Kueng, 
 A partial derandomization of phase lift using spherical designs, 
 {\sl J. Fourier Anal. Appl} {\bf 21}, 229 (2015).

 \bibitem{BHM18} J. Bae, B. C. Hiesmayr, and D. McNulty,
 Linking Entanglement Detection and State Tomography via Quantum $2$-Designs,  
{\sl New J. Phys.} {\bf 21}, 013012 (2019).
 
\bibitem{GA16} M. A. Graydon, and D. M. Appleby, 
Quantum Conical Designs, 
{\sl J. Phys. A}  {\bf 49}, 085301 (2016).

\bibitem{Szymusiak}  S. Brandsen, M. Dall’Arno, and A. Szymusiak,
Communication capacity of mixed quantum $t$-designs,
{\sl Phys. Rev. A} {\bf 94}, 022335 (2016).

\bibitem{SZ84} P. D. Seymour and T. Zaslavsky,
Averaging sets: A generalization of mean values and spherical designs,
{\sl Advanc. Math.} {\bf  52}, 213 (1984).

\bibitem{La11} J. Lawrence, 
Entanglement patterns in mutually unbiased basis sets, 
{\sl Phys. Rev. A} {\bf 84}, 022338 (2011).

 \bibitem{H82} S. G. Hoggar, $t$-Designs in Projective Spaces, 
{\sl Europ. J. Combinatorics} {\bf 3}, 233 (1982). 

\bibitem{Pe94} R. Penrose, 
{\sl Shadows of the Mind},
 Oxford University Press, Oxford, 1994.

\bibitem{ZP93} J. Zimba  and R. Penrose,
On Bell non-locality without probabilities: more curious geometry. 
{\sl Stud. Hist. Phil. Sci.} {\bf  24}, 697 (1993).

\bibitem{MA99} J. E. Massad and P. K. Aravind, 
The Penrose dodecahedron revisited, 
{\sl American Journal of Physics} {\bf 67}, 631 (1999).

\bibitem{Zi06} J. Zimba,  Anticoherent spin states via the
Majorana representation,
{\sl Elect. J. Theor. Phys.} {\bf 3},  143 (2006).


\bibitem{BQTSLSKB15} N. Bent, H. Qassim, A. A. Tahir, D. Sych,
  G. Leuchs, L. L. Sanchez-Soto, E. Karimi, and R. W. Boyd, 
  Experimental
  Realization of Quantum Tomography of Photonic Qudits via Symmetric
  Informationally Complete Positive Operator-Valued Measures, 
{\sl  Phys. Rev. X} {\bf 5}, 041006 (2015).

 \bibitem{ZS01} K. {\.Z}yczkowski and H.-J. Sommers, 
Induced measures in the space of mixed quantum states, 
{\sl J. Phys. A} {\bf 34}, 7111 (2001).

\bibitem{We74} L. Welch,  Lower bounds on the maximum cross correlation of signals,
 {\sl  IEEE Transactions Inform Theor.} {\bf 20}, 397 (1974).


\bibitem{Zh15} H. Zhu,
 Super-symmetric informationally complete measurements,
 {\sl Annals Phys.} {\bf  362}, 311 (2015).

 
 \bibitem{RTH15} J. {\v R}eh{\'a}{\v c}ek, Y. S. Teo, and Z. Hradil,
 On determining which quantum measurement performs better for state estimation,
 {\sl Phys. Rev. A} {\bf 92}, 012108 (2015).
 

\bibitem{SHBAH12} Ch. Spengler, M. Huber, St. Brierley, Th. Adaktylos and
B.C. Hiesmayr,
 Entanglement detection via mutually unbiased bases,
 {\sl Phys. Rev. A} {\bf 86}, 022311 (2012).

\bibitem{CSW18} D. Chru{\' s}ci{\' n}ski, G. Sarbicki and F. Wudarski, Entanglement witnesses from mutually unbiased bases,  	{\sl Phys. Rev. A} {\bf 97}, 032318 (2018).   

 \bibitem{MBPKIN14} A. Miranowicz, K. Bartkiewicz,  J. Pe{\v r}ina Jr., 
 M. Koashi, N. Imoto, and F. Nori,
 Optimal two-qubit tomography based on local and global measurements:
 Maximal robustness against errors as described by condition numbers,
 {\sl Phys. Rev. A} {\bf 90}, 062123 (2014). 

 
 \bibitem{BCLM16} K. Bartkiewicz,  A. {\v C}ernoch, K. Lemr, and A. Miranowicz,
  Priority Choice Experimental Two-Qubit Tomography:
   Measuring One by One All Elements of Density Matrices,
 {\sl Scient. Rep.} {\bf 6}, 19610 (2016).
 

\bibitem{WA17} M. Waegell and P. K. Aravind, The  Penrose dodecahedron and the Witting polytope are identical in $\mathbb{C}P^3$, 
{\sl Phys. Lett. A} {\bf 381}, 1853 (2017).


\end{thebibliography}
\end{document}